\documentclass[11pt]{article}
\usepackage{graphicx}
\usepackage{verbatim,hyperref,multirow,cite,authblk}
\usepackage{amssymb, color}
\begin{document}
\title{Machine Learning and Materials Informatics: \\ Recent Applications and Prospects}
\date{}

\author[1]{Rampi Ramprasad\footnote{Corresponding author; email: rampi.ramprasad@uconn.edu}}
\author[1]{Rohit Batra}
\author[2,3]{Ghanshyam Pilania}
\author[1,4]{Arun Mannodi-Kanakkithodi}
\author[1]{Chiho Kim}
\affil[1]{Department of Materials Science \& Engineering and Institute of Materials
	 Science, University of Connecticut, 97 North Eagleville Rd., Unit 3136, 
	 Storrs, CT 06269-3136, USA}
\affil[2]{Fritz-Haber-Institut der Max-Planck-Gesellschaft, Faradayweg 4-6, D-14195 Berlin, Germany}
\affil[3]{Materials Science and Technology Division, Los Alamos National Laboratory, Los Alamos, New Mexico 87545, USA}
\affil[4]{Center for Nanoscale Materials, Argonne National Laboratory, 9700 S. Cass Ave., Lemont, IL 60439, USA}
\maketitle
\begin{abstract}
Propelled partly by the Materials Genome Initiative, and partly by the algorithmic developments and the resounding successes of data-driven efforts in other domains, informatics strategies are beginning to take shape within materials science. These approaches lead to surrogate machine learning models that enable rapid predictions based purely on past data rather than by direct experimentation or by computations/simulations in which fundamental equations are explicitly solved. Data-centric informatics methods are becoming useful to determine material properties that are hard to measure or compute using traditional methods\textemdash due to the cost, time or effort involved\textemdash but for which reliable data either already exists or can be generated for at least a subset of the critical cases. Predictions are typically interpolative, involving fingerprinting a material numerically first, and then following a mapping (established via a learning algorithm) between the fingerprint and the property of interest. Fingerprints may be of many types and scales, as dictated by the application domain and needs. Predictions may also be extrapolative\textemdash extending into new materials spaces\textemdash provided prediction uncertainties are properly taken into account. This article attempts to provide an overview of some of the recent successful data-driven ``materials informatics" strategies undertaken in the last decade, and identifies some challenges the community is facing and those that should be overcome in the near future.
\end{abstract}

\newpage
\section*{Overarching Perspectives}
When a new situation is encountered, cognitive systems (including humans) have a natural tendency to make decisions based on past similar encounters. When the new situation is distinctly different from those encountered in the past, errors in judgment may occur and lessons may be learned. The sum total of such past scenarios, decisions made and the lessons learned may be viewed collectively as ``experience", ``intuition" or even as ``common sense". Ideally, depending on the intrinsic capability of the cognitive system, its ability to make decisions should progressively improve as the richness of scenarios encountered increases.

In recent decades, the artificial intelligence (AI) and statistics communities have made these seemingly vague notions quantitative and mathematically precise\cite{1_Gopnik2017-em,2_Jordan2015-cs}. These efforts have resulted in practical machines that learn from past experiences (or ``examples"). Classic exemplars of such \textit{machine learning} approaches include facial, fingerprint or object recognition systems, machines that can play sophisticated games such as chess, Go or poker, and automation systems such as in robotics or self-driving cars. In each of these cases, a large dataset of past examples is required, e.g., images and their identities, configuration of pieces in a board game and the best moves, and scenarios encountered while driving and the best actions.

On the surface, it may appear as though the ``data-driven" approach for determining the best decision or answer when a new situation or problem is encountered is radically different from approaches based on fundamental science in which predictions are made by solving equations that govern the pertinent phenomena. But viewed differently, isn't the scientific process itself\textemdash which begins with observations, followed by intuition, then construction of a quantitative theory that explains the observations, and subsequently, refinement of the theory based on new observations\textemdash the ultimate culmination of such data-driven inquiries?

\begin{figure}[h]
  \begin{center}
  \includegraphics[width=1\linewidth]{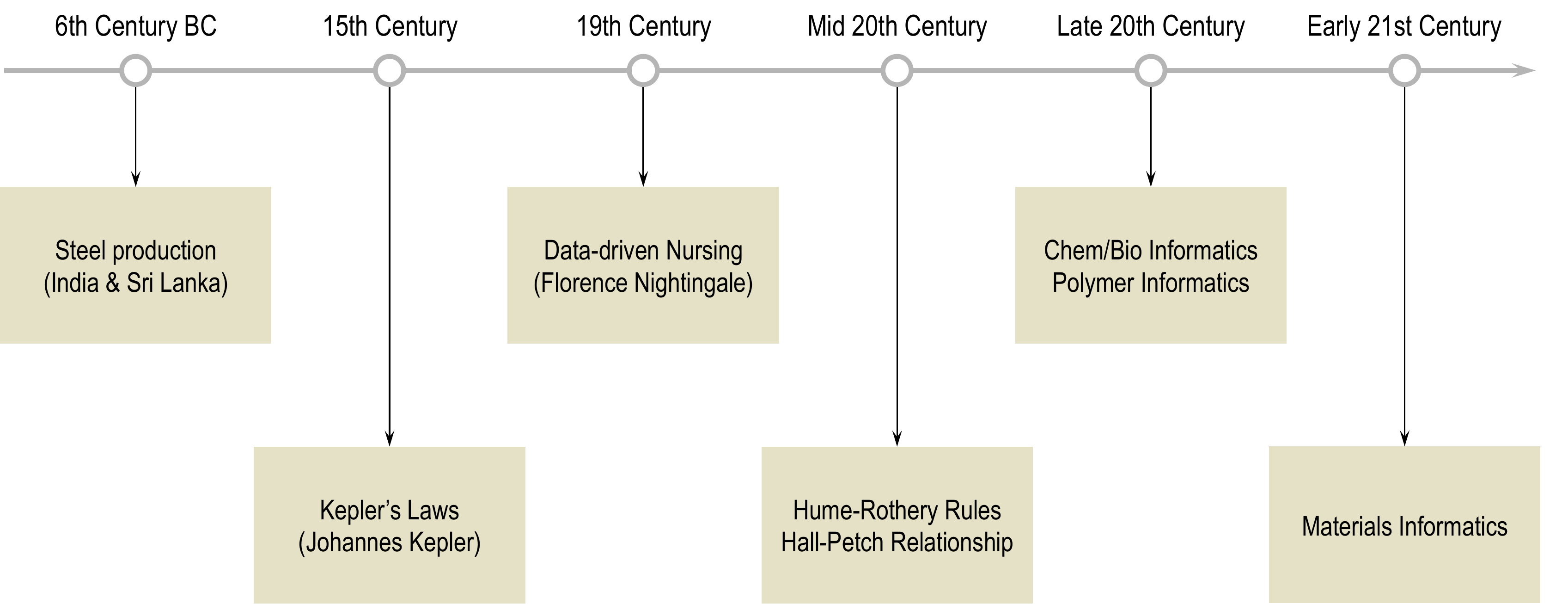}
  \caption{Some classic historical examples of data-driven science and engineering efforts.}
  \label{fig:timeline}
  \end{center}
\end{figure}
For instance, consider how the ancient people from India and Sri Lanka figured out, through persistent tinkering, the alloying elements to add to iron to impede its tendency to rust, using only their experience and creativity\cite{3_Srinivasan2004-jo,4_Ward2008-ey} (and little ``steel science", which arose from this empiricism much later)\textemdash an early example of the reality and power of ``chemical intuition." Or, more recently, over the last century, consider the enormously practical Hume-Rothery rules to determine the solubility tendency of one metal in another\cite{5_Jwc1961-qh}, the Hall-Petch studies that have led to empirical relationships between grain sizes and mechanical strength (not just for metals but for ceramics as well)\cite{6_Hall1951-bt,7_Petch1986-km}, and the group contribution approach to predict complex properties of organic and polymeric materials based just on the identity of the chemical structure\cite{8_Van_Krevelen2009-eu}, all of which arose from data-driven pursuits (although they were not called as such), and later rationalized using physical principles. It would thus be fair to say that data\textemdash either directly or indirectly\textemdash drives the creation of both complex fundamental and simple empirical scientific theories. Figure 1 charts the timeline for some classic historical and diverse examples of data-driven efforts.

In more modern times, in the last decade or so, thanks to the implicit or explicit acceptance of the above notions, the ``data-driven", ``machine learning" or ``materials informatics" paradigms (with these terms used interchangeably by the community) are rapidly becoming an essential part of the materials research portfolio\cite{9_Mueller2016-ch,10_Ward2017-av,11_Green2017-cj,12_Hattrick-Simpers2016-kg}. The availability of robust and trustworthy \textit{in silico} simulation methods and systematic synthesis and characterization capabilities, although time-consuming and sometimes expensive, provide a pathway to generate at least a subset of the required critical data in a targeted and organized manner (e.g., via ``high-throughput" experiments or computations). Mining or learning from this or other \emph{reliable} extant data can lead to the recognition of previously unknown correlations between properties, and the discovery of qualitative and quantitative rules\textemdash also referred to as \textit{surrogate models}\textemdash that can be used to predict material properties orders of magnitude faster and cheaper, and with reduced human effort than required by the benchmark simulation or experimental methods utilized to create the data in the first place.

With excitement and opportunities come challenges. Questions constantly arise as to what sort of materials science problems are most appropriate for, or can benefit most from, a data-driven approach. A satisfactory understanding of this aspect is essential before one makes a decision on using machine learning methods for their problem of interest. Perhaps the most dangerous aspect of data-driven approaches is the unwitting application of machine learning models to cases that fall outside the domain of prior data. A rich and largely uncharted area of inquiry is to recognize when such a scenario ensues, and to be able to quantify the uncertainties of the machine learning predictions especially when models veer out-of-domain. Solutions for handling these perilous situations may open up pathways for \textit{adaptive} learning models that can progressively improve in quality through systematic infusion of new data\textemdash an aspect critical to the further burgeoning of machine learning within the hard sciences.

This article attempts to provide an overview of some of the recent successful data-driven materials research strategies undertaken in the last decade, and identifies challenges that the community is facing and those that should be overcome in the near future.

\section*{Elements of Machine Learning (Within Materials Science)}
Regardless of the specific problem under study, a prerequisite for machine learning is the existence of past data. Thus, either clean, curated and reliable data corresponding to the problem under study should already be available, or an effort has to be put in place upfront for the creation of such data. An example dataset may be an enumeration of a variety of materials that fall within a well-defined chemical class of interest and a relevant measured or computed property of those materials (see Figure 2a). Within the machine learning parlance, the former, i.e., the material, is referred to as ``input", and the latter, i.e., the property of interest, is referred to as the ``target" or ``output." \textbf{A learning problem (Figure 2b) is then defined as follows: Given a \{materials $\rightarrow$ property\} dataset, what is the best estimate of the property for a new material not in the original dataset?} Provided that there are sufficient examples, i.e., that the dataset is sufficiently large, and provided that the new material falls within the same chemo-structural class as the materials in the original dataset, we expect that it should be possible to make such an estimate. Ideally, uncertainties in the prediction should also be reported, which can give a sense of whether the new case is within or outside the domain of the original dataset.

\begin{figure}[h]
  \begin{center}
  \includegraphics[width=1\linewidth]{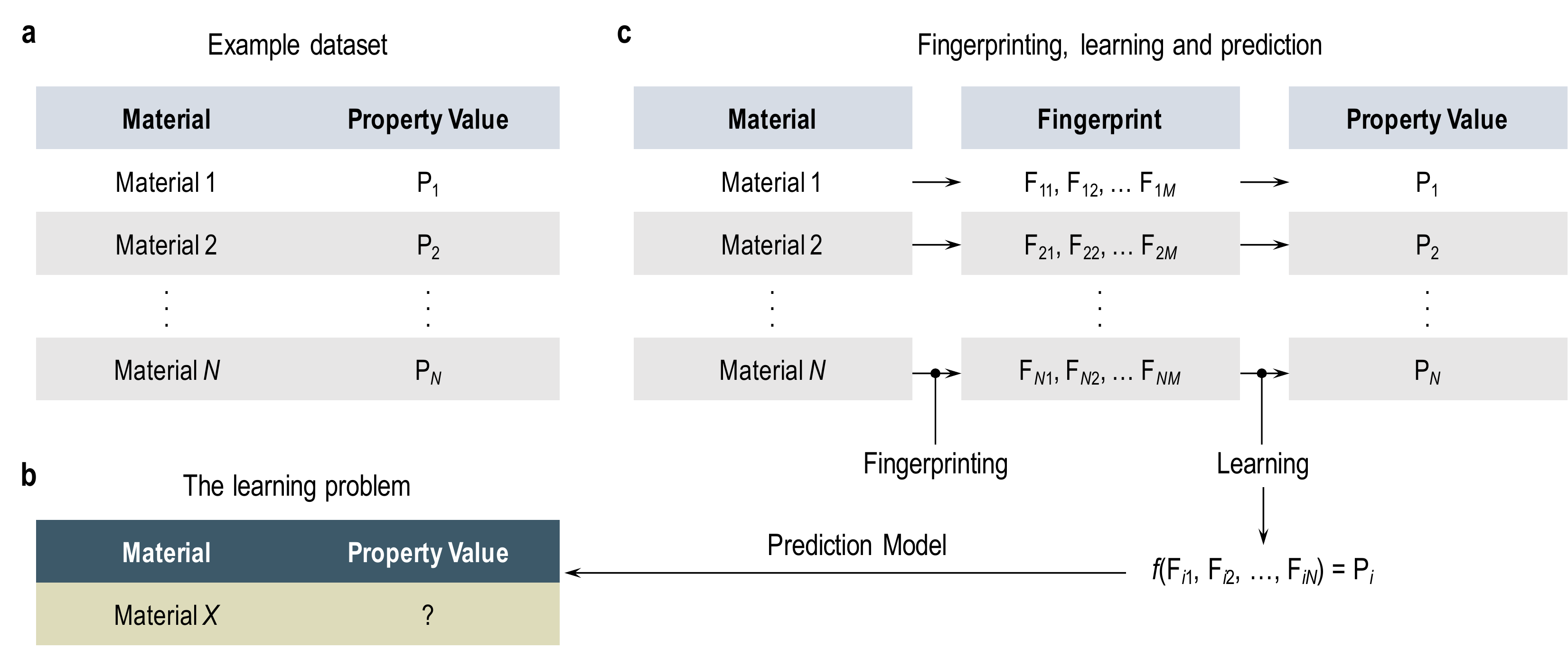}
  \caption{The key elements of machine learning in materials science. (a) Schematic view of an example dataset, (b) statement of the learning problem, and (c) creation of a surrogate prediction model via the fingerprinting and learning steps. $N$ and $M$ are, respectively, the number of training examples and the number of fingerprint (or descriptor or feature) components.}
  \label{fig:ml_scheme}
  \end{center}
\end{figure}

All data-driven strategies that attempt to address the problem posed above are composed of two distinct steps, both aimed at satisfying the need for quantitative predictions. The first step is to represent numerically the various input cases (or materials) in the dataset. At the end of this step, each input case would have been reduced to a string of numbers (or ``fingerprints"; see Figure 2c). This is such an enormously important step,  requiring significant expertise and knowledge of the materials class and the application, i.e., ``domain expertise", that we devote a separate Section to its discussion below.

The second step establishes a mapping between the fingerprinted input and the target property, and is entirely numerical in nature, largely devoid of the need for domain knowledge. Both the fingerprinting and mapping/learning steps are schematically illustrated in Figure 2. Several algorithms, ranging from elementary (e.g., linear regression) to highly sophisticated (kernel ridge regression, decision trees, deep neural networks), are available to establish this mapping and the creation of surrogate prediction models\cite{13_Bishop2006-kp,14_Theodoridis2015-eu,15_Hastie2013-gy}. While some algorithms provide actual functional forms that relate input to output (e.g., regression based schemes), others do not (e.g., decision trees). In the above discussion, it was implicitly assumed that the target property is a continuous quantity (e.g., bulk modulus, band gap, melting temperature, etc.). Problems can also involve discrete targets (e.g., crystal structure, specific structural motifs, etc.), which are referred to as classification problems.

Throughout the above learning process, it is typical (and essential) to adhere to rigorous statistical practices. Central to this is the notion of cross-validation and testing on unseen data, which attempt to ensure that a learning model developed based on the original dataset can truly handle a new case without falling prey to the perils of ``overfitting"\cite{9_Mueller2016-ch,15_Hastie2013-gy}.

Machine learning should be viewed as the sum total of the organized creation of the initial dataset, the fingerprinting and learning steps, and a necessary subsequent step (discussed at the end of this article) of progressive and targeted new data infusion, ultimately leading to an expert recommendation system that can continuously and adaptively improve.

\section*{Hierarchy of Fingerprints or Descriptors}
We now elaborate on what is perhaps \textit{the} most important component of the machine learning paradigm, the one that deals with the numerical representation of the input cases or materials. A numerical representation is essential to make the prediction scheme quantitative (i.e., moving it away from the ``vague" notions alluded to in the first paragraph of this article). The choice of the numerical representation can be effectively accomplished \textit{only} with adequate knowledge of the problem and goals (i.e., domain expertise or experience), and typically proceeds in an iterative manner by duly considering aspects of the material that the target property may be correlated with. Given that the numerical representation serves as the proxy for the real material, it is also referred to as the fingerprint of the material or its \textit{descriptors} (in machine learning parlance, it is also referred to as the feature vector).

Depending on the problem under study and the accuracy requirements of the predictions, the fingerprint can be defined at varying levels of granularity. For instance, if the goal is to obtain a high-level understanding of the factors underlying a complex phenomenon\textemdash such as the mechanical or electrical strength of materials, catalytic activity, etc.\textemdash and prediction accuracy is less critical, then the fingerprint may be defined at a gross level, e.g., in terms of the general attributes of the atoms the material is made up of, other potentially relevant properties (e.g., the band gap) or higher-level structural features (e.g., typical grain size). On the other hand, if the goal is to predict specific properties at a reasonable level of accuracy across a wide materials chemical space\textemdash such as the dielectric constant of an insulator or the glass transition temperature of a polymer\textemdash the fingerprint may have to include information pertaining to key atomic-level structural fragments that may control these properties. If extreme (chemical) accuracy in predictions is demanded\textemdash such as total energies and atomic forces, precise identification of structural features, space groups or phases\textemdash the fingerprint has to be fine enough so that it is able to encode details of atomic-level structural information with sub-Angstrom-scale resolution. Several examples of learning based on this hierarchy of fingerprints or descriptors are provided in subsequent Sections.

The general rule of thumb is that finer the fingerprint, greater is the expected accuracy, and more laborious, more data-intensive and less conceptual is the learning framework. A corollary to the last point is that rapid coarse-level initial screening of materials should generally be targeted using coarser fingerprints.

Regardless of the specific choice of representation, the fingerprints should also be invariant to certain transformations. Consider the facial recognition scenario. The numerical representation of a face should not depend on the actual placement location of the face in an image, nor should it matter whether the face has been rotated or enlarged with respect to the examples the machine has seen before. Likewise, the representation of a material should be invariant to the rigid translation or rotation of the material. If the representation is fine enough that it includes atomic position information, permutation of like atoms should not alter the fingerprint. These invariance properties are easy to incorporate in coarser fingerprint definitions but non-trivial in fine-level descriptors. Furthermore, ensuring that a fingerprint contains all the relevant components (and only the relevant components) for a given problem requires careful analysis, for example, using unsupervised learning algorithms\cite{9_Mueller2016-ch,15_Hastie2013-gy}. For these reasons, construction of a fingerprint for a problem at hand is not always straightforward or obvious.

\section*{Examples of learning based on gross-level property-based descriptors}
Two historic efforts in which gross-level descriptors were utilized to create surrogate models (although they were not couched under those terms) have lead to the Hume-Rothery rules\cite{5_Jwc1961-qh} and Hall-Petch relationships\cite{6_Hall1951-bt,7_Petch1986-km} (Figure 1). The former effort may be viewed as a classification exercise in which the target is to determine whether a mixture of two metals will form a solid solution; the gross-level descriptors considered were the atomic sizes, crystal structures, electronegativities and oxidation states of the two metal elements involved. In the latter example, the strength of a polycrystalline material is the target property, which was successfully related to the average grain size; specifically a linear relationship was found between the strength and the reciprocal of the square root of the average grain size. While a careful manual analysis of data gathered from experimentation was key to developing such rules in the past, modern machine learning and data mining approaches provide powerful pathways for such knowledge discovery, especially when the dependencies are multivariate and highly nonlinear.

\begin{figure}
	\begin{center}
		\includegraphics[width=1\linewidth]{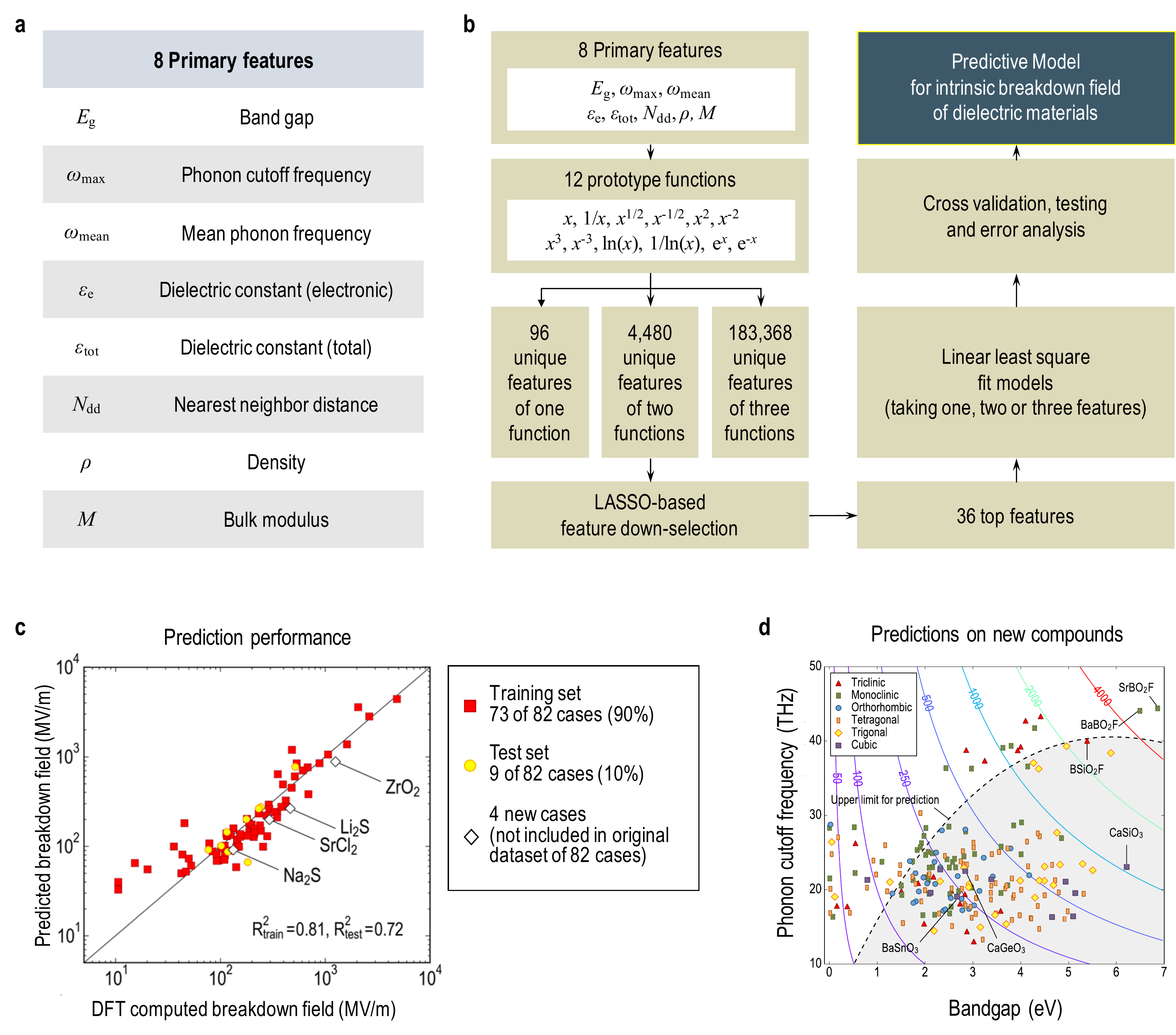}
		\caption{Building phenomenological models for the prediction of the intrinsic electrical breakdown field of insulators. (a) Primary features expected to correlate to the intrinsic breakdown field; (b) Creation of compound features,  down-selection of a subset of critical compound features using LASSO and predictive model building; (c) Final phenomenological model performance versus DFT computations for the binary octet dataset [adapted with permission from \cite{20_Kim2016-uh}. Copyright (2017) American Chemical Society]; and (d) Application of the model for the identification of new breakdown resistant perovskite type materials (contours represent predicted breakdown field in MV/m and the model's prediction domain is depicted in gray color) [adapted with permission from \cite{21_Kim2016-fc}. Copyright (2017) American Chemical Society].}
		\label{fig:level1}
	\end{center}
\end{figure}

To identify potential nonlinear multivariate relationships efficiently, one may start from a moderate number of potentially relevant primary descriptors (e.g., electronegativity, $E$, ionic radius, $R$, etc.), and create millions or even billions of compound descriptors by forming algebraic combinations of the primary descriptors (e.g., $E/R^2$, $R\log(E)$, etc.); see Figures 3a and 3b. This large space of nonlinear mathematical functions needs to be ``searched" for a subset that is highly correlated with the target property. Dedicated methodological approaches to accomplish such a task have emerged from recent work in genetic programing\cite{16_Schmidt2009-pn}, compressed sensing\cite{17_Ghiringhelli2015-bb,18_Ghiringhelli2017-wt}, and information science\cite{19_Lookman2015-xa}.

One such approach\textemdash based on the least absolute shrinkage and selection operator (LASSO)\textemdash was recently demonstrated to be highly effective for determining key physical factors that control a complex phenomenon through identification of simple empirical relationships\cite{17_Ghiringhelli2015-bb,18_Ghiringhelli2017-wt}. An example of such complex behavior is the tendency of insulators to fail when subjected to extreme electric fields\cite{20_Kim2016-uh,21_Kim2016-fc}. The critical field at which this failure occurs in a defect-free material\textemdash referred to as the \emph{intrinsic} electrical breakdown field\textemdash is related to the balance between energy gained by charge carriers from the electric field to the energy lost due to collisions with phonons. The intrinsic breakdown field may be computed from first principles by treatment of electron-phonon interactions, but this computation process is enormously laborious. Recently, the breakdown field was computed from first principles using density functional theory (DFT) for a benchmark set of 82 binary octet insulators\cite{20_Kim2016-uh}. This dataset included alkali metal halides, transition metal halides, alkaline earth metal chalcogenides, transition metal oxides, and group III, II-VI, I-VII semiconductors. After validating the theoretical results by comparing against available experimental data, this dataset was used to build simple predictive phenomenological  surrogate models of dielectric breakdown using LASSO as well as other advanced machine learning schemes. The general flow of the LASSO-based procedure, starting from the primary descriptors considered (Figure 3a), is charted in Figure 3b. The trained and validated surrogate models were able to reveal key correlations and analytical relationships between the breakdown field and other easily accessible material properties such as the band gap and the phonon cutoff frequency. Figure 3c shows the agreement between such a discovered analytical relationship and the DFT results (spanning 3 orders of magnitude) for the benchmark dataset of 82 insulators, as well as for 4 new ones that were not included in the original training dataset.

The phenomenological model was later employed to systematically screen and identify perovskite compounds with high breakdown strength. The purely machine learning based screening revealed that boron-containing compounds are of particular interest, some of which were predicted to exhibit remarkable intrinsic breakdown strength of $\sim$1 GV/m (see Figure 3d). These predictions were subsequently confirmed using first principles computations\cite{21_Kim2016-fc}.

The LASSO-based and related schemes have also been shown to be enormously effective at predicting the preferred crystal structures of materials. In a pioneering study that utilized the LASSO-based approach, Ghiringelli and co-workers were able to classify binary octet insulators into tendencies for the formation of rock salt versus zinc blende structures\cite{17_Ghiringhelli2015-bb,18_Ghiringhelli2017-wt,22_Goldsmith2017-vg}. More recently, Bialon and co-workers\cite{23_Bialon2016-gf} aimed to classify 64 different prototypical crystal structures formed by A$_{\rm x}$B$_{\rm y}$ type compounds, where A and B are \textit{sp}-block and transition metal elements, respectively. After searching over a set of $1.7\times10^5$ non-linear descriptors formed by physically meaningful functions of primary coarse-level descriptors such as band-filling, atomic volume, and different electronegativity scales of the \textit{sp} and \textit{d} elements, the authors were able to find a set of 3 optimal descriptors. A three-dimensional structure-map\textemdash built on the identified descriptor set\textemdash was used to classify 2,105 experimentally known training examples available from the Pearson's Crystal Database\cite{24_noauthor_2008-qb} with an 86\% probability of predicting the correct crystal structure. Likewise, Oliynyk and co-workers recently used a set of elemental descriptors to train a machine-learning model, built on a random forest algorithm\cite{25_Oliynyk2016-bd}, with an aim to accelerate the search for Heusler compounds. After training the model on available crystallographic data from Pearson's Crystal Database\cite{24_noauthor_2008-qb} and the ASM Alloy Phase Diagram Database\cite{26_noauthor_undated-rh} the model was used to evaluate the probabilities at which compounds with the formula AB$_2$C will adopt Heusler structures. This approach was exceptionally successful in distinguishing between Heusler and non-Heusler compounds (with a true positive rate of 94\%), including the prediction of unknown compounds and flagging erroneously assigned entries in the literature and in crystallographic databases. As a proof of concept, 12 novel predicted candidates (Gallides with formulae MRu$_2$Ga and RuM$_2$Ga, where M = Ti, V, Cr, Mn, Fe and Co) were synthesized and confirmed to be Heusler compounds.

Yet another application of the gross-level descriptors relate to the prediction of the band gap of insulators\cite{27_Dey2014-sq,28_Ward2016-ia,29_Lee2016-lr,30_Pilania2016-ig,31_Pilania2017-kq}. Rajan and co-workers\cite{27_Dey2014-sq} have used experimentally available band gaps of ABC$_2$ chalcopyrite compounds to train regression models with electronegativity, atomic number, melting point, pseudopotential radii, and the valence for each of the A, B and C elements as features. Just using the gross-level elemental features, the developed machine learning models were able to predict the experimental band gaps with moderate accuracy. In a different study, Pilania and co-workers\cite{30_Pilania2016-ig} used a database consisting of computed band gaps of $\sim$1,300 AA'BB'O$_6$ type double perovskites to train a kernel ridge regression (KRR) machine learning model, a scheme that allows for nonlinear relationships based on measures of (dis)similarity between fingerprints, for efficient predictions of the band gaps. A set of descriptors with increasing complexity was identified by searching across a large portion of the feature space using LASSO, with more than $\sim$1.2 million compound descriptors created from primary elemental features such as electronegativities, ionization potentials, electronic energy levels and valence orbital radii of the constituent atomic species. One of the most important chemical insights that emerged from this effort was that the band gap in the double perovskites is primarily controlled (and therefore effectively learned) by the lowest occupied energy levels of the A-site elements and electronegativities of the B-site elements.

Other successful attempts of using gross-level descriptors include the creation of surrogate models for the estimation of formation enthalpies\cite{32_Faber2016-ab,33_Meredig2014-vi,34_Deml2016-dt}, free energies\cite{35_Legrain2017-kp}, defect energetics\cite{36_Medasani2016-ss}, melting temperatures\cite{37_Seko2014-ov,38_Pilania2015-yw}, mechanical properties\cite{39_Chatterjee2007-yz,40_De_Jong2016-ki,41_Aryal2014-ab}, thermal conductivity\cite{42_Seko2015-qe}, catalytic activity\cite{43_Li2017-oq,44_Hong2016-lt}, and radiation damage resistance\cite{45_Pilania2017-ci}. Efforts are also underway for the identification of novel shape memory alloys\cite{46_Xue2016-us}, improved piezoelectrics\cite{47_Xue2016-gu}, MAX phases\cite{47_Xue2016-gu,48_Ashton2016-fe}, novel perovskite\cite{49_Pilania2016-at} and double perovskite halides\cite{32_Faber2016-ab,49_Pilania2016-at}, CO$_2$ capture materials\cite{50_Fernandez2014-bq} and potential candidates for water splitting\cite{51_Emery2016-cx}.

Emerging materials informatics tools also offer tremendous potential and new avenues for mining for structure-property-processing linkages from aggregated and curated materials datasets\cite{52_Kalidindi2016-ss}. While a large fraction of such efforts in the current literature has considered relatively simple definitions of the material that included mainly the overall chemical composition of the material, Kalidindi and co-workers\cite{53_Brough2017-zp,54_Kalidindi2015-tk,55_Gupta2015-fs,56_Brough2017-gh} have recently proposed a new materials data science framework known as Materials Knowledge Systems (MKS)\cite{57_Panchal2013-bf,58_Brough2017-vy} that explicitly accounts for the complex hierarchical material structure in terms of \textit{n}-point spatial correlations (also frequently referred to as \textit{n}-point statistics). Further adopting the \textit{n}-point statistics as measures to quantify materials microstructure, a flexible computational framework has been developed to customize toolsets to understand structure-property-processing linkages in materials science\cite{59_Kalidindi2012-xw}.

\section*{Examples of learning based on molecular fragment-level descriptors}
The next in the hierarchy of descriptor types are those that encode finer details than those captured by the gross-level properties. Within this class, materials are described in terms of the basic building blocks they are made of. The origins of ``block-level" or ``molecular fragment" based descriptors can be traced back to cheminformatics, which is a field of theoretical chemistry that deals with correlating properties such as biological activity, physio-chemical properties and reactivity with molecular structure and fragments\cite{60_Adamson1974-uy,61_Adamson1974-jp,62_Judson2009-zq}, leading up to what is today referred to as quantitative structure activity/property relationships (QSAR/QSPR).

\begin{figure}
	\begin{center}
		\includegraphics[width=1\linewidth]{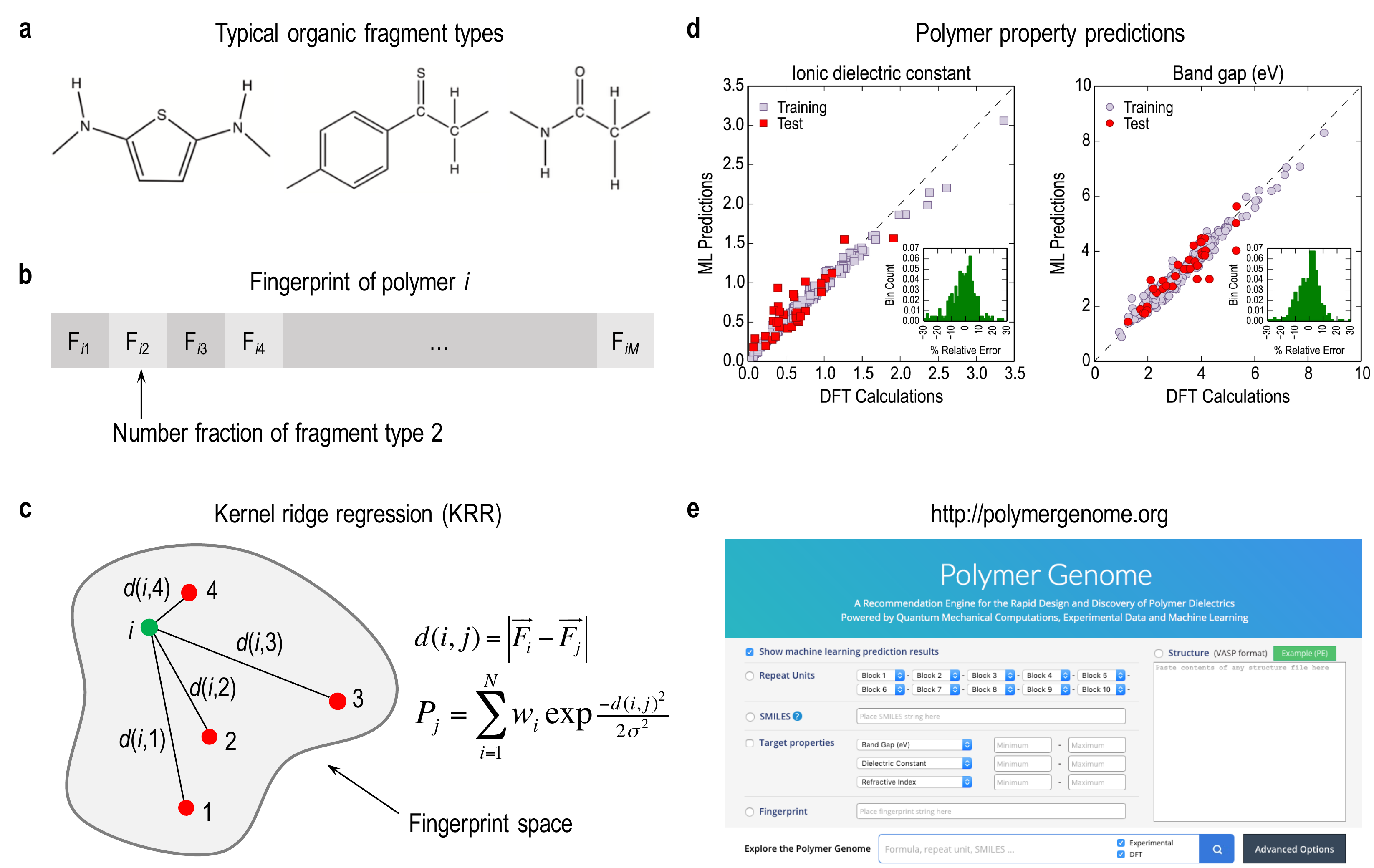}
		\caption{Learning polymer properties using fragment-level fingerprints. (a) Typical fragments that can be used for the case of organic molecules, crystals or polymers; (b) Schematic of organic polymer fingerprint construction; (c) Schematic of the kernel ridge regression (KRR) scheme showing the example cases in fingerprint ($F$) space. The distance, $d(i,j)$, between the point (in fingerprint space) corresponding to a new case, $j$, and each of the training example cases, $i$, is used to predict the property, $P_j$, of case $j$; (d) Surrogate machine learning (ML) model predictions versus DFT results for key dielectric polymer properties \cite{68_Mannodi-Kanakkithodi2016-vc}; (e) Snapshot of the \emph{Polymer Genome} online application for polymer property prediction.}
		\label{fig:level2}
	\end{center}
\end{figure}

Within materials science, specifically, within polymer science, the notions underlying QSAR/QSPR ultimately led to the successful group contribution methods\cite{8_Van_Krevelen2009-eu}. Van Krevelen and co-workers studied the properties of polymers and discovered that they were strongly correlated to the chemical structure (i.e., nature of the polymer repeat unit, end groups, etc.) and the molecular weight distribution. They observed that polymer properties such as glass transition temperature, solubility parameter and bulk modulus (which were, and still are, difficult to compute using traditional computational methods) were correlated with the presence of chemical groups and combinations of different groups in the repeat unit. Based on a purely data-driven approach, they developed an ``atomic group contribution method" to express various properties as a linear weighted sum of the contribution (called atomic group parameter) from every atomic group that constituted the repeat unit. These groups could be units like CH$_2$, C$_6$H$_4$, CH$_2$-CO, etc., that make up the polymer. It was also noticed that factors such as the presence of aromatic rings, long side chains and cis/trans conformations influence the properties, prompting their introduction into the group additivity scheme. For instance, a CH$_2$ group attached to an aromatic ring would have a different atomic group parameter than a CH$_2$ group attached to an aliphatic group. In this fashion, nearly all the important contributing factors were taken into account, and linear empirical relationships were devised for thermal, elastic and other polymer properties. However, widespread usage of these surrogate models is still restricted because (1) the definition of atomic groups is somewhat \textit{ad hoc}, and (2) the target properties are assumed to be linearly related to the group parameters.

Modern data-driven methods have significantly improved on these earlier ideas with regards to both issues mentioned above. Recently, in order to enable the accelerated discovery of polymer dielectrics\cite{63_Huan2016-kk,64_Mannodi-Kanakkithodi2016-zv,65_Treich2017-xn,66_Huan2016-kl,67_Sharma2014-re,114_Lorenzini}, hundreds of polymers built from a chemically allowed combination of 7 possible basic units, namely, CH$_2$, CO, CS, O, NH, C$_6$H$_4$ and C$_4$H$_2$S, were considered, inclusive of van der Waals interactions\cite{115_Chunsheng}, and a set of properties relevant for dielectric applications, namely, the dielectric constant and band gap, were computed using DFT\cite{63_Huan2016-kk,68_Mannodi-Kanakkithodi2016-vc}. These polymers were then fingerprinted by keeping track of the occurrence of a fixed set of molecular fragments in the polymers in terms of their number fractions\cite{68_Mannodi-Kanakkithodi2016-vc,69_Pilania2013-ec}. A particular molecular fragment could be a triplet of contiguous blocks such as -NH-CO-CH$_2$- (or, at a finer level, a triplet of contiguous atoms, such as C$_4$-O$_2$-C$_3$ or C$_3$-N$_3$-H$_1$, where X$_{n}$ represents an $n$-fold coordinated X atom)\cite{70_T_D_Huan_A_Mannodi-Kanakkithodi_R_Ramprasad2015-rg,71_A_Mannodi-Kanakkithodi_T_D_Huan_R_Ramprasad_undated-xm}. All possible triplets were considered (some examples are shown in Figure 4a), and the corresponding number fractions in a specific order formed the fingerprint of a particular polymer (see Figure 4b). This procedure provides a uniform and seamless pathway to represent all polymers within this class, and the procedure can be indefinitely generalized by considering higher order fragments (i.e., quadruples, quintuples, etc., of atom types). Furthermore, relationships between the fingerprint and properties have been established using the KRR learning algorithm; a schematic of how this algorithm works is shown in Figure 4c. The capability of this scheme for dielectric constant and band gap predictions is portrayed in Figure 4d. These predictive tools are available online (Figure 4e) and are constantly being updated\cite{72_noauthor_undated-cb}. The power of such modern data-driven molecular fragment-based learning approaches (like its group contribution predecessor) lies in the realization that \textit{any} type of property related to the molecular structure\textemdash whether computable using DFT (e.g., band gap, dielectric constant) or measurable experimentally (e.g., glass transition temperature, dielectric loss)\textemdash can be \textit{learned} and predicted.

The molecular fragment-based representation is not restricted to polymeric materials. Novel compositions of A$_{\rm x}$B$_{\rm y}$O$_{\rm z}$ ternary oxides and their most probable crystal structures have been predicted using a probabilistic model built on an experimental crystal structure database\cite{73_Hautier2010-ou}. The descriptors used in this study are a combination of the type of crystal structure (spinel, olivine, etc.) and the composition information, i.e. the elements that constitute the compound. Likewise, surrogate machine learning models have been developed for predicting the formation energies of A$_{\rm x}$B$_{\rm y}$O$_{\rm z}$ ternary compounds using only compositional information as descriptors, trained on a dataset of 15,000 compounds from the Inorganic Crystal Structure Database\cite{33_Meredig2014-vi}. Using this approach, 4,500 new stable materials have been discovered. Finally, surrogate models have been developed for predicting the formation energies of elpasolite crystals with the general formula A$_2$BCD$_6$, based mainly on compositional information. The descriptors used take into account the periodic table row and column of elements A, B, C and D that constitute the compound (although this fingerprint could have been classified as a gross-level one, we choose to place this example in the present Section as the prototypical structure of the elpasolite was implicitly assumed in this work and fingerprint). Important correlations and trends were revealed between atom types and the energies; for example, it was found that the preferred element for the D site is F, and that for the A and B sites are late group II elements\cite{32_Faber2016-ab}.

\section*{Examples of learning based on sub-Angstrom-level descriptors}
We now turn to representing materials at the finest possible scale, such that the fingerprint captures precise details of atomic configurations with high fidelity. Such a representation is useful in many scenarios. For instance, one may attempt to connect this fine-scale fingerprint directly with the corresponding total potential energy with chemical accuracy, or with structural phases/motifs (e.g., crystal structure or the presence/absence of a stacking fault). The former capability can lead to purely data-driven accelerated atomistic computational methods, and the latter to refined and efficient on-the-fly characterization schemes.

``Chemical accuracy" specifically refers to potential energy and reaction enthalpy predictions with errors of less than 1 kcal/mol, and atomic force predictions (the input quantity for molecular dynamics, or MD, simulations) with errors of less than 0.05 eV/\AA. Chemical accuracy is key to enable reliable MD simulations (or for precise identification of the appropriate structural phases or motifs), and is only possible with fine-level fingerprints that offer sufficiently high configurational resolution, more than those in the examples encountered thus far.

The last decade has seen spectacular activity and successes in the general area of data-driven atomistic computations. All modern atomistic computations use either some form of quantum mechanical scheme (e.g., DFT) or a suitably parameterized semi-empirical method to predict the properties of materials, given just the atomic configuration. Quantum mechanical methods are versatile, i.e., they can be used to study any material, in principle. However, they are computationally demanding, as complex differential equations governing the behavior of electrons are solved for every given atomic configuration. Systems involving at most about 1,000 atoms can be simulated routinely in a practical setting today. In contrast, semi-empirical methods use prior knowledge about interatomic interactions under known conditions and utilize parameterized analytical equations to determine properties such as the total potential energies, atomic forces, etc. These semi-empirical \textit{force fields} are several orders of magnitude faster than quantum mechanical methods, and are the choice today for routinely simulating systems containing millions to billions of atoms, as well as the dynamical evolution of systems at nonzero temperatures (using the MD method) at timescales of nanoseconds to milliseconds. However, a major drawback of traditional semi-empirical force fields is that they lack versatility, i.e., they are not transferable to situations or materials for which the original functional forms and parameterizations don't apply.

Machine learning is rapidly bridging the chasm between the two extremes of quantum mechanical and semi-empirical methods, and has offered surrogate models that combine the best of both worlds. Rather than resort to specific functional forms and parameterizations adopted in semi-empirical methods (the aspects that restrict their versatility), machine learning methods use an \{atomic configuration $\rightarrow$ property\} dataset, carefully prepared, e.g., using DFT, to make interpolative predictions of the property of a new configuration at speeds several orders of magnitude faster than DFT. Any material for which adequate reference DFT computations may be performed ahead of time can be handled using such a machine learning scheme. Thus, the lack of versatility issue of traditional semi-empirical approach and the time-intensive nature of quantum mechanical calculations are simultaneously addressed, while also preserving quantum mechanical and chemical accuracy.

The primary challenge though has been the creation of suitable fine-level fingerprinting schemes for materials, as these fingerprints are required to be strictly invariant with respect to arbitrary translations, rotations, and exchange of like atoms, in addition to being continuous and differentiable (i.e., ``smooth") with respect to small variations in atomic positions. Several candidates, including those based on symmetry functions\cite{74_Behler2007-ig,75_Behler2008-tn,76_Behler2014-zf}, bispectra of neighborhood atomic densities\cite{77_Bartok2010-zz}, Coulomb matrices (and its variants)\cite{78_Rupp2012-in,79_Chmiela2017-pp}, smooth overlap of atomic positions (SOAP)\cite{80_Bartok2013-tj,81_Szlachta2014-dp,82_Bartok2015-ap,83_Deringer2017-sq}, and others\cite{84_Jindal2017-ov,85_Thompson_undated-ka}, have been proposed. Most fingerprinting approaches use sophisticated versions of distribution functions (the simplest one being the radial distribution function) to represent the distribution of atoms around a reference atom as qualitatively captured in Figure 5a. The Coulomb matrix is an exception, which elegantly represents a molecule, with the dimensionality of the matrix being equal to the total number of atoms in the molecule. Although questions have arisen with respect to smoothness considerations and whether the representation is under/over-determined (depending on whether the eigenspectrum or the entire matrix is used as the fingerprint)\cite{80_Bartok2013-tj}, this approach has been shown to be able to predict various molecular properties accurately\cite{79_Chmiela2017-pp}.

Figure 5b also shows a general schema typically used in the construction of machine learning force fields, to be used in MD simulations. Numerous learning algorithms\textemdash ranging from neural networks, KRR, Gaussian process regression (GPR), etc.\textemdash have been utilized to accurately map the fingerprints to various materials properties of interest. A variety of fingerprinting schemes as well as learning schemes that lead up to force fields have been recently reviewed\cite{9_Mueller2016-ch,80_Bartok2013-tj,86_Rupp2015-si}. One of the most successful and widespread machine learning force field schemes to date is the one by Behler and co-workers\cite{74_Behler2007-ig}, which uses symmetry function fingerprints mapped to the total potential energy using a neural network. Several applications have been studied, including surface diffusion, liquids, phase equilibria in bulk materials, etc. This approach is also quite versatile in that multiple elements can be considered. Bispectra based fingerprints combined with GPR learning schemes have lead to Gaussian approximation potentials (GAP)\cite{74_Behler2007-ig,77_Bartok2010-zz}, which have also been demonstrated to provide chemical accuracy, versatility and efficiency.

A new development within the area of machine learning force fields is to learn and predict the atomic forces directly\cite{87_Li2015-nv,88_Botu2015-jb,89_Glielmo2017-tr,90_Botu2015-aa,91_Botu2017-vk,92_Botu2017-kr}; the total potential energy is determined through appropriate integration of the forces along a reaction coordinate or MD trajectory\cite{92_Botu2017-kr}. These approaches are inspired by Feynman'€™s original idea that it should be possible to predict atomic forces given just the atomic configuration, without going through the agency of the total potential energy\cite{93_Feynman1939-nm}. An added attraction of this perspective is that the atomic force can be uniquely assigned to an individual atom, while the potential energy is a global property of the entire system (partitioning the potential energy to atomic contributions does not have a formal basis). Mapping atomic fingerprints to purely atomic properties can thus lead to powerful and accurate prescriptions. Figure 5c, for instance, compares the atomic forces at the core of an edge dislocation in Al, predicted using a machine learning force prediction recipe called AGNI, with the DFT forces for the same atomic configuration. Also shown are forces predicted using the embedded atom method (EAM), a popular classical force field, for the same configuration. EAM tends to severely under-predict large forces while the machine learning scheme predicts forces with high fidelity (neither EAM nor the machine learning force field were explicitly trained on dislocation data). This general behavior is consistent with recent detailed comparisons of EAM with machine learning force fields\cite{94_Bianchini2016-qe}.

\begin{figure}
\centering
\includegraphics[width=1\linewidth]{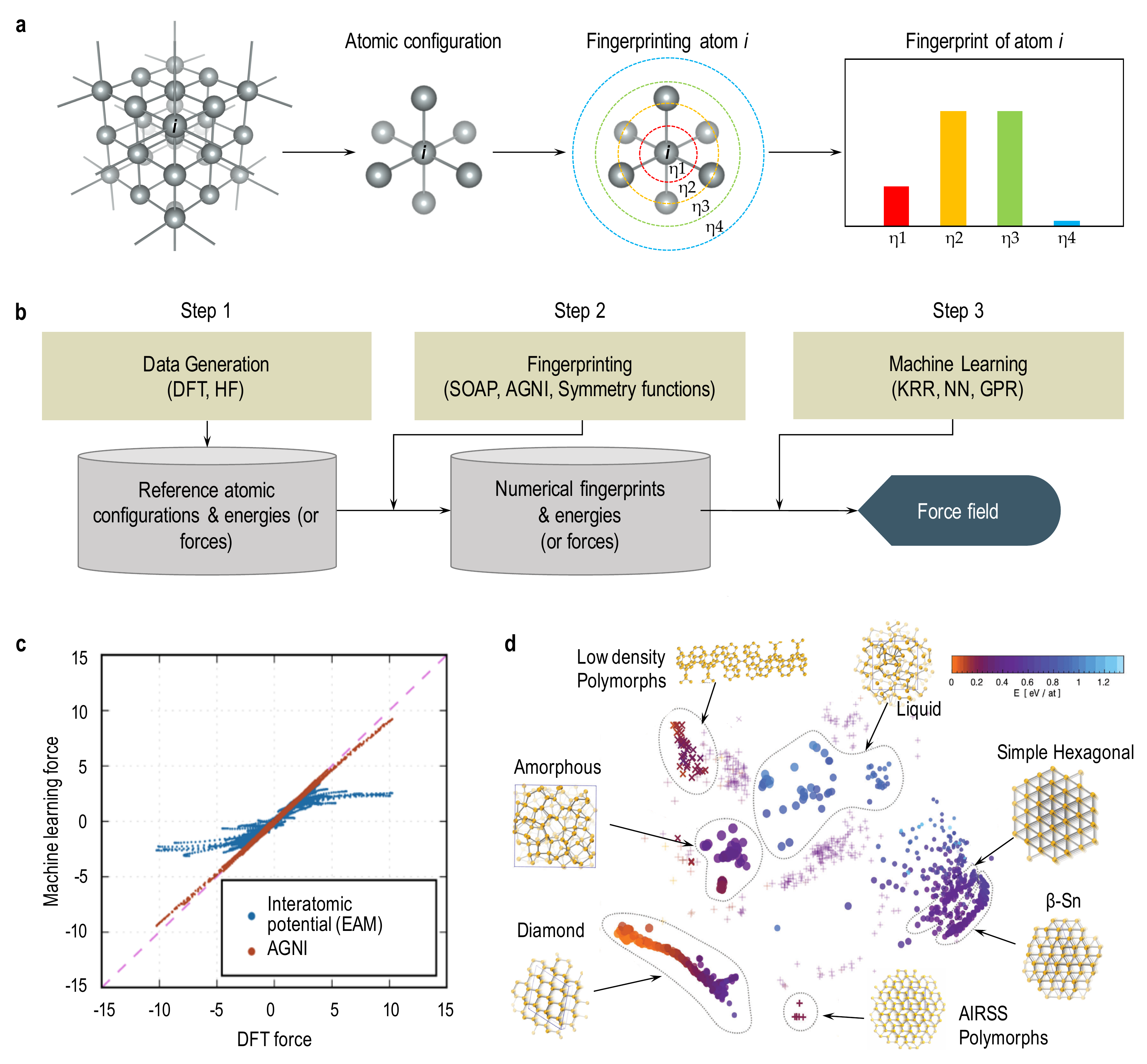}
\caption{Learning from fine-level fingerprints. (a) A schematic portrayal of the sub-Angstrom-level atomic environment fingerprinting scheme adopted by Behler and co-workers. $\eta_j$s denote the widths of Gaussians, indexed by $j$, placed at the reference atom $i$ whose environment needs to be fingerprinted. The histograms in the right represent the integrated number of atoms within each Gaussian sphere; (b) Schematic of a typical workflow for the construction of machine learning force fields; (c) Prediction of atomic forces in the neighborhood of an edge dislocation in bulk Al using the atomic force-learning scheme AGNI and the embedded atom method (EAM), and comparison with the corresponding DFT results [adapted with permission from \cite{92_Botu2017-kr}. Copyright (2017) American Chemical Society]; (d) Classifying atomic environments in Si using the SOAP fingerprinting scheme and the "Sketch Map" program for dimensionality reduction [adapted with permission from \cite{95_De2016-al}. Copyright (2017) Royal Society of Chemistry].}
\label{fig:level3}
\end{figure}

Another notable application of fine-level fingerprints has been in the use of  the electronic charge density itself as the representation to learn various properties\cite{69_Pilania2013-ec} or density functionals\cite{96_Snyder2012-aq,97_Snyder2013-bp,98_Snyder2015-ad}, thus going to the very heart of DFT. While these efforts are in a state of infancy\textemdash as they have dealt with mainly toy problems and learning the kinetic energy functional\textemdash such efforts have great promise as they attempt to integrate machine learning methods within DFT (all other DFT-related informatics efforts so far have utilized machine learning external to DFT).

Fine-level fingerprints have also been used to characterize structure in various settings. Within a general crystallographic structure refinement problem, one has to estimate the structural parameters of a system, i.e., the unit cell parameters (a, b, c, $\alpha$, $\beta$, and $\gamma$) that best fit measured X-ray diffraction (XRD) data. Using a Bayesian learning approach and a Markov chain Monte Carlo (MCMC) algorithm to sample multiple combinations of possible structural parameters for the case of Si, Fancher and co-workers\cite{99_Fancher2016-db} not only accurately determined the estimates of the structural parameters, but also quantified the associated uncertainty (thus going beyond the conventional Rietveld refinement method).

Unsupervised learning using fine-level fingerprints (and clustering based on these fingerprints) has led to the classification of materials based on their phases or structural characteristics\cite{11_Green2017-cj,12_Hattrick-Simpers2016-kg}. Using the XRD spectrum itself as the fingerprint, high-throughput XRD measurements for various compositional spreads\cite{11_Green2017-cj,12_Hattrick-Simpers2016-kg,100_Kusne2014-dt,101_Kusne2015-or,102_noauthor_2016-al,103_Bunn2016-wb} have been used to automate the creation of phase diagrams. Essentially, features of the XRD spectra are used to distinguish between phases of a material as a function of composition. Likewise, on the computational side, the SOAP fingerprints have been effectively used to distinguish between different allotropes of materials, as well as different motifs that emerge during the course of a MD simulation (see Figure 5d for an example)\cite{95_De2016-al}.

\section*{Critical steps going forward}
\subsection*{Quantifying the uncertainties of predictions}
Given that machine learning predictions are inherently statistical in nature, uncertainties must be expected in the predictions. Moreover, predictions are typically and ideally \textit{interpolative} between data points corresponding to previously seen data. To what extent a new case for which a prediction needs to be made falls in or out of the domain of the original dataset (i.e., to what extent the predictions are interpolative or extrapolative) may be quantified using the predicted uncertainty. While strategies are available to prescribe prediction uncertainties, these ideas have been explored only to a limited extent within materials science\cite{46_Xue2016-us,104_Lookman2017-fj}. Bayesian methods (e.g., Gaussian process regression)\cite{15_Hastie2013-gy} provide a natural pathway for estimating the uncertainty of the prediction in addition to the prediction itself. This approach assumes that a Gaussian distribution of models fit the available data, and thus a distribution of predictions may be made. The mean and variance of these predictions\textemdash the natural outcomes of Bayesian approaches\textemdash are the most likely predicted value and the uncertainty of the prediction, respectively, within the spectrum of models and the fingerprint considered. Other methods may also be utilized to estimate uncertainties, but at significant added cost. A straightforward and versatile scheme is bootstrapping\cite{105_Felsenstein2008-gk}, in which different (but small) subsets of the data are randomly excluded, and several prediction models are developed based on these closely related but modified datasets. The mean and variance of the predictions from these bootstrapped models provide the property value and expected uncertainty. Essentially, this approach attempts to probe how sensitive the model is with respect to slight ``perturbations" to the dataset. Another related methodology is to explicitly consider a variety of closely related models, e.g., neural networks or decision trees with slightly different architectures, and to use the distribution of predictions to estimate uncertainty\cite{76_Behler2014-zf}.

\subsection*{Adaptive learning and design}
Uncertainty quantification has a second important benefit. It can be used to continuously and progressively improve a prediction model, i.e., render it a \textit{truly learning} model. Ideally, the learning model should adaptively and iteratively improve by asking questions such as \textit{``what should be the next new material system to consider or include in the training set that would lead to an improvement of the model or the material?"} This may be accomplished by balancing the tradeoffs between "exploration" and "exploitation"\cite{104_Lookman2017-fj,106_Powell2012-hb}. That is, at any given stage of an iterative learning process, a number of new candidates may be predicted to have certain properties with uncertainties. The tradeoff is between exploiting the results by choosing to perform the next computation (or experiment) on the material predicted to have the optimal target property or further improving the model through exploration by performing the calculation (or experiment) on a material where the predictions have the largest uncertainties. This can be done rigorously by adopting well-established information theoretic selector frameworks such as the knowledge gradient\cite{107_Powell2010-cp,108_Ryzhov2012-aw}. In the initial stages of the iterative process, it is desired to ``explore and learn" the property landscape. As the machine learning predictions improve and the associated uncertainties shrink, the adaptive design scheme allows one to gradually move away from exploration towards exploitation. Such an approach, schematically portrayed in Figure 6a, enables one to systematically expand the training data towards a target chemical space where materials with desired functionality are expected to reside.

\begin{figure}[t]
	\centering
	\includegraphics[width=1\linewidth]{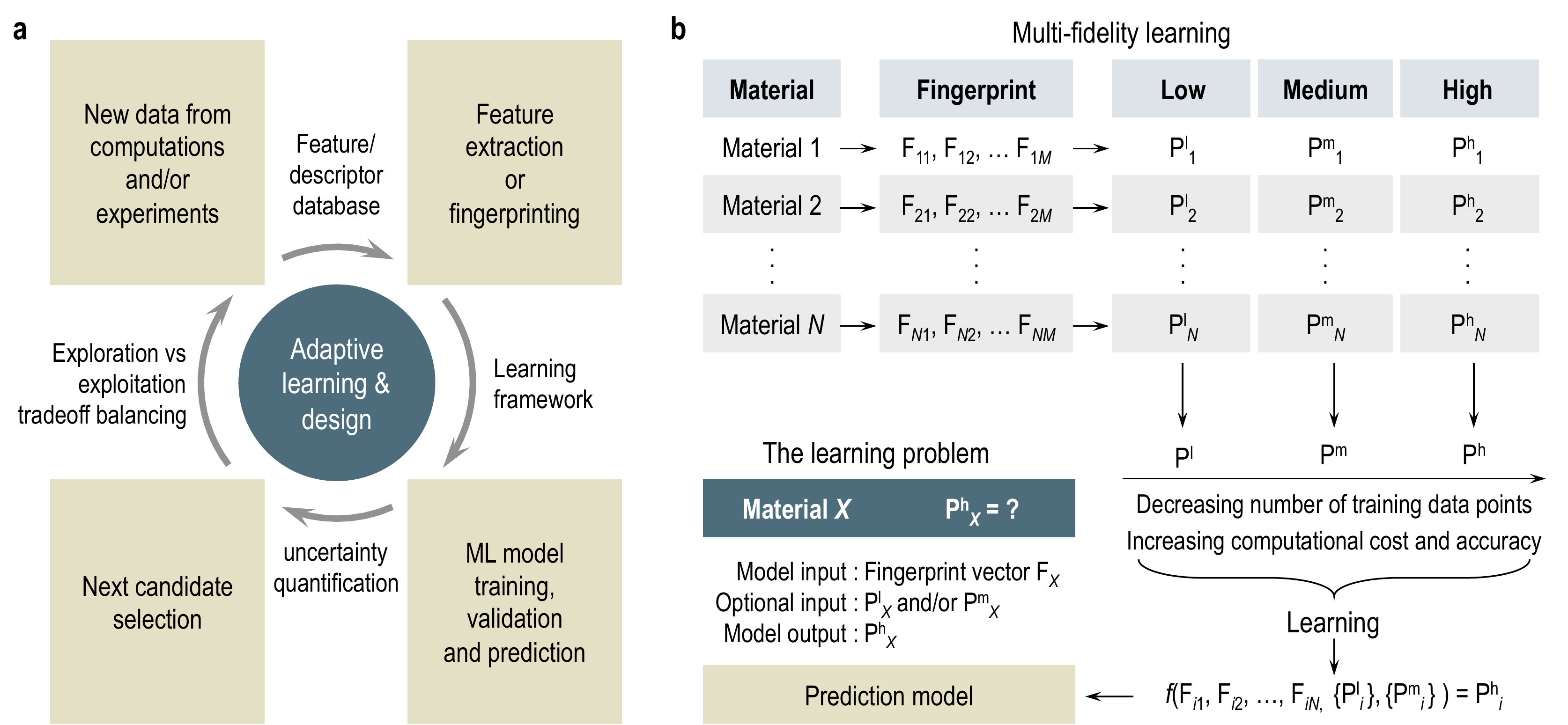}
	\caption{(a) Schematic illustration of adaptive design via balanced exploration and exploitation enabled by uncertainty quantification; (b) An example dataset used in a multi-fidelity learning setting involving target properties obtained at various levels of fidelity and expense, and the statement of the multi-fidelity learning problem.}
	\label{fig:new_design}
\end{figure}

Some of the first examples of using adaptive design for targeted materials discovery include identification of shape memory alloys with low thermal hysteresis\cite{46_Xue2016-us} and accelerated search for BaTiO$_3$-based piezoelectrics with optimized morphotropic phase boundary\cite{47_Xue2016-gu}. In the first example, Xue and co-workers\cite{46_Xue2016-us} employed the aforementioned adaptive design framework to find NiTi-based shape memory alloys that may display low thermal hysteresis. Starting from a limited number of 22 training examples and going through the iterative process 9 times, 36 predicted compositions were synthesized and tested from a potential space of $\sim$800,000 compound possibilities. It was shown that 14 out of these 36 new compounds were better (i.e., had a smaller thermal hysteresis) than any of the 22 compounds in the original data set. The second successful demonstration of the adaptive design approach combined informatics and Landau-€"Devonshire theory to guide experiments in the design of lead-free piezoelectrics\cite{47_Xue2016-gu}. Guided by predictions from the machine learning model, an optimized solid solution, (Ba$_{0.5}$Ca$_{0.5}$)TiO$_3$-Ba(Ti$_{0.7}$Zr$_{0.3}$)O$_3$, with piezoelectric properties was synthesized and characterized to show better temperature reliability than other BaTiO$_3$-based piezoelectrics in the initial training data.

\subsection*{Other algorithms}
The materials science community is just beginning to explore and utilize the plethora of available information theoretic algorithms to mine and learn from data. The usage of an algorithm is driven largely by need, as it should. One such need is to be able to learn and predict vectorial quantities. Examples include functions, such as the electronic or vibrational density of states (which are functions of energy or frequency). Although, the target property in these cases may be viewed as a set of scalar quantities at each energy or frequency (for a given structure) to be learned and predicted independently, it is desirable to learn and predict the entire function simultaneously. This is because the value of the function at a particular energy or frequency is correlated to the function values at other energy or frequency values. Properly learning the function of interest requires machine learning algorithms that can handle vectorial outputs. Such algorithms are indeed available\cite{109_Micchelli2005-lj,110_Alvarez2012-wa}, and if exploited can lead to prediction schemes of the electronic structure for new configurations of atoms. Another class of examples where vector learning is appropriate includes cases where the target property is truly a vector (e.g., atomic force) or a tensor (e.g., stress). In these cases, the vector or tensor transforms in a particular way as the material itself is transformed, e.g., if it is rotated (in the examples of functions discussed above, the vectors, i.e., the functions, are invariant to any unitary transformation of the material). These truly vectorial or tensorial target property cases will thus have to be handled with care, as has been done recently using vector learning and covariant kernels\cite{89_Glielmo2017-tr}.

Another algorithm that is beginning to show value within material science falls under multi-fidelity learning\cite{111_Forrester2007-so}. This learning method can be used when a property of interest can be computed at several levels of fidelities, exhibiting a natural hierarchy in both computational cost and accuracy. A good materials science example is the band gap of insulators computed at an inexpensive lower level of theory, e.g., using a semilocal electronic exchange-correlation functional (the low-fidelity value), and the band gap computed using an more accurate, but expensive, approach, e.g., using a hybrid exchange-correlation functional (the high-fidelity value). A naive approach in such a scenario can be to use a low-fidelity property value as a feature in a machine learning model to predict the corresponding higher fidelity value. However, using low-fidelity estimates as features strictly requires the low-fidelity data for all materials for which predictions are to be made using the trained model. This can be particularly challenging and extremely computationally demanding when faced with a combinatorial problem that targets exploring vast chemical and configurational spaces. A multi-fidelity co-kriging framework, on the other hand, can seamlessly combine inputs from two or more levels of fidelities to make accurate predictions of the target property for the highest fidelity. Such an approach, schematically represented in Figure 6b, requires high-fidelity training data only on a subset of compounds for which low-fidelity training data is available. More importantly, the trained model can make efficient highest-fidelity predictions even in the absence of the low-fidelity data for the prediction set compounds. While multi-fidelity learning is routinely used in several fields to address computationally challenging engineering design problems\cite{111_Forrester2007-so,112_Perdikaris2015-yt}, it is only beginning to find applications in materials informatics\cite{31_Pilania2017-kq}.

Finally, machine learning algorithms may also lead to strategies for making the so-called ``inverse design" of materials possible. Inverse design refers to the paradigm whereby one seeks to identify materials that satisfy a target set of desired properties (in this parlance, the ``forward" process refers to predicting the properties of a given material)\cite{113_Dudiy2006-ge}. Within the machine learning context, although the backward process of going from a desired set of properties to the appropriate fingerprints is straightforward, the process of inverting the fingerprint to actual physically and chemically meaningful materials continues to be a major hurdle. Two strategies that are adopted to achieve inverse design within the context of machine learning involves either inverting the desired properties to only fingerprints that correspond to physically realizable materials (through imposition of constraints that fingerprint components are required to satisfy)\cite{70_T_D_Huan_A_Mannodi-Kanakkithodi_R_Ramprasad2015-rg,113_Dudiy2006-ge}, or adopting schemes such as the genetic algorithm or simulated annealing to determine iteratively a population of materials that meet the given target property requirements\cite{68_Mannodi-Kanakkithodi2016-vc,70_T_D_Huan_A_Mannodi-Kanakkithodi_R_Ramprasad2015-rg}. Despite these developments, true inverse design continues to remain a challenge (although materials design through adaptive learning discussed above appears to have somewhat mitigated this challenge).

\subsection*{Decisions on when to use machine learning}
Perhaps the most important question that plagues new researchers eager to use data-driven methods is whether their problem lends itself to such methods. Needless to say, the existence of past \textit{reliable} data, or efforts devoted to its generation for at least a subset of the critical cases in a uniform and controlled manner, is a prerequisite for the adoption of machine learning. Even so, the question is the appropriateness of machine learning for the problem at hand. Ideally, data-driven methods should be aimed at (1) properties very difficult or expensive to compute or measure using traditional methods, (2) phenomena that are complex enough (or nondeterministic) that there is no hope for a direct solution based on solving fundamental equations, or (3) phenomena whose governing equations are not (yet) known, providing a rationale for the creation of surrogate models. Such scenarios are replete in the social, cognitive and biological sciences, explaining the pervasive applications of data-driven methods in such domains. Materials science examples ideal for studies using machine learning methods include properties such as the glass transition temperature of polymers, dielectric loss of polycrystalline materials over a wide frequency and temperature range, mechanical strength of composites, failure time of engineering materials (e.g., due to electrical, mechanical or thermal stresses), friction coefficient of materials, etc., all of which involve the inherent complexity of materials, i.e., their polycrystalline or amorphous nature, multi-scale geometric architectures, the presence of defects of various scales and types, and so on. Machine learning may also be used to eliminate redundancies underlying repetitive but expensive operations, especially when interpolations in high-dimensional spaces are required, such as when properties across enormous chemical and/or configurational spaces are desired. An example of the latter scenario, i.e., an immense configurational space, is encountered in first principles molecular dynamics simulations, when atomic forces are evaluated repetitively (using expensive quantum mechanical schemes) for myriads of very similar atomic configurations. The area of machine learning force fields has burgeoned to meet this need. Yet another setting where large chemical and configurational spaces are encountered is the emerging domain of high-throughput materials characterization, where on-the-fly predictions are required to avoid data accumulation bottlenecks. Although materials informatics efforts so far have largely focused on model problems and the validation of the general notion of data-driven discovery, active efforts are beginning to emerge that focus on complex real-world materials applications, strategies to handle situations inaccessible to traditional materials computations, and the creation of adaptive prediction frameworks (through adequate uncertainty quantification) that build efficiencies within rational materials design efforts.

\section*{Acknowledgements}
The authors acknowledge financial support from several grants from the Office of Naval Research that allowed them to explore many applications of machine learning within materials science. Several engaging discussions with Kenny Lipkowitz, Huan Tran and Venkatesh Botu are gratefully acknowledged. GP acknowledges the Alexander von Humboldt Foundation.

\section*{Author contributions}
RR lead the creation of the manuscript, with critical contributions on various sections and graphics by GP, RB, AMK and CK. All authors participated in the writing of the manuscript.


\begin{thebibliography}{100}
	\expandafter\ifx\csname url\endcsname\relax
	\def\url#1{\texttt{#1}}\fi
	\expandafter\ifx\csname urlprefix\endcsname\relax\def\urlprefix{URL }\fi
	\providecommand{\bibinfo}[2]{#2}
	\providecommand{\eprint}[2][]{\url{#2}}
	
	\bibitem{1_Gopnik2017-em}
	\bibinfo{author}{Gopnik, A.}
	\newblock \bibinfo{title}{Making {AI} more human}.
	\newblock \emph{\bibinfo{journal}{Sci. Am.}} \textbf{\bibinfo{volume}{316}},
	\bibinfo{pages}{60--65} (\bibinfo{year}{2017}).
	
	\bibitem{2_Jordan2015-cs}
	\bibinfo{author}{Jordan, M.~I.} \& \bibinfo{author}{Mitchell, T.~M.}
	\newblock \bibinfo{title}{Machine learning: Trends, perspectives, and
		prospects}.
	\newblock \emph{\bibinfo{journal}{Science}} \textbf{\bibinfo{volume}{349}},
	\bibinfo{pages}{255--260} (\bibinfo{year}{2015}).
	
	\bibitem{3_Srinivasan2004-jo}
	\bibinfo{author}{Srinivasan, S.} \& \bibinfo{author}{Ranganathan, S.}
	\newblock \emph{\bibinfo{title}{India's Legendary Wootz Steel: An Advanced
			Material of the Ancient World}} (\bibinfo{publisher}{National Institute of
		advanced studies}, \bibinfo{year}{2004}).
	
	\bibitem{4_Ward2008-ey}
	\bibinfo{author}{Ward, G. W.~R.}
	\newblock \emph{\bibinfo{title}{The Grove Encyclopedia of Materials and
			Techniques in Art}} (\bibinfo{publisher}{Oxford University Press},
	\bibinfo{year}{2008}).
	
	\bibitem{5_Jwc1961-qh}
	\bibinfo{author}{{J.W.C.}}
	\newblock \bibinfo{title}{Atomic theory for students of metallurgy}.
	\newblock \emph{\bibinfo{journal}{Journal of the Less Common Metals}}
	\textbf{\bibinfo{volume}{3}}, \bibinfo{pages}{264} (\bibinfo{year}{1961}).
	
	\bibitem{6_Hall1951-bt}
	\bibinfo{author}{Hall, E.~O.}
	\newblock \bibinfo{title}{The deformation and ageing of mild steel: {III}
		discussion of results}.
	\newblock \emph{\bibinfo{journal}{Proc. Phys. Soc. B}}
	\textbf{\bibinfo{volume}{64}}, \bibinfo{pages}{747--753}
	(\bibinfo{year}{1951}).
	
	\bibitem{7_Petch1986-km}
	\bibinfo{author}{Petch, N.~J.}
	\newblock \bibinfo{title}{The influence of grain boundary carbide and grain
		size on the cleavage strength and impact transition temperature of steel}.
	\newblock \emph{\bibinfo{journal}{Acta Metall.}} \textbf{\bibinfo{volume}{34}},
	\bibinfo{pages}{1387--1393} (\bibinfo{year}{1986}).
	
	\bibitem{8_Van_Krevelen2009-eu}
	\bibinfo{author}{Van~Krevelen, D.~W.} \& \bibinfo{author}{Te~Nijenhuis, K.}
	\newblock \emph{\bibinfo{title}{Properties of Polymers: Their Correlation with
			Chemical Structure; their Numerical Estimation and Prediction from Additive
			Group Contributions}} (\bibinfo{publisher}{Elsevier}, \bibinfo{year}{2009}).
	
	\bibitem{9_Mueller2016-ch}
	\bibinfo{author}{Mueller, T.}, \bibinfo{author}{Kusne, A.~G.} \&
	\bibinfo{author}{Ramprasad, R.}
	\newblock \bibinfo{title}{Machine learning in materials science}.
	\newblock In \emph{\bibinfo{booktitle}{Reviews in Computational Chemistry}},
	\bibinfo{pages}{186--273} (\bibinfo{publisher}{John Wiley \& Sons, Inc},
	\bibinfo{year}{2016}).
	
	\bibitem{10_Ward2017-av}
	\bibinfo{author}{Ward, L.} \& \bibinfo{author}{Wolverton, C.}
	\newblock \bibinfo{title}{Atomistic calculations and materials informatics: A
		review}.
	\newblock \emph{\bibinfo{journal}{Curr. Opin. Solid State Mater. Sci.}}
	\textbf{\bibinfo{volume}{21}}, \bibinfo{pages}{167--176}
	(\bibinfo{year}{2017}).
	
	\bibitem{11_Green2017-cj}
	\bibinfo{author}{Green, M.~L.} \emph{et~al.}
	\newblock \bibinfo{title}{Fulfilling the promise of the materials genome
		initiative with high-throughput experimental methodologies}.
	\newblock \emph{\bibinfo{journal}{Applied Physics Reviews}}
	\textbf{\bibinfo{volume}{4}}, \bibinfo{pages}{011105} (\bibinfo{year}{2017}).
	
	\bibitem{12_Hattrick-Simpers2016-kg}
	\bibinfo{author}{Hattrick-Simpers, J.~R.}, \bibinfo{author}{Gregoire, J.~M.} \&
	\bibinfo{author}{Kusne, A.~G.}
	\newblock \bibinfo{title}{Perspective: Composition--structure--property mapping
		in high-throughput experiments: Turning data into knowledge}.
	\newblock \emph{\bibinfo{journal}{APL Materials}} \textbf{\bibinfo{volume}{4}},
	\bibinfo{pages}{053211} (\bibinfo{year}{2016}).
	
	\bibitem{13_Bishop2006-kp}
	\bibinfo{author}{Bishop, C.~M.}
	\newblock \emph{\bibinfo{title}{Pattern Recognition and Machine Learning}}
	(\bibinfo{publisher}{Springer}, \bibinfo{year}{2006}).
	
	\bibitem{14_Theodoridis2015-eu}
	\bibinfo{author}{Theodoridis, S.}
	\newblock \emph{\bibinfo{title}{Machine Learning: A Bayesian and Optimization
			Perspective}} (\bibinfo{publisher}{Academic Press}, \bibinfo{year}{2015}).
	
	\bibitem{15_Hastie2013-gy}
	\bibinfo{author}{Hastie, T.}, \bibinfo{author}{Tibshirani, R.} \&
	\bibinfo{author}{Friedman, J.}
	\newblock \emph{\bibinfo{title}{The Elements of Statistical Learning: Data
			Mining, Inference, and Prediction}} (\bibinfo{publisher}{Springer Science \&
		Business Media}, \bibinfo{year}{2013}).
	
	\bibitem{20_Kim2016-uh}
	\bibinfo{author}{Kim, C.}, \bibinfo{author}{Pilania, G.} \&
	\bibinfo{author}{Ramprasad, R.}
	\newblock \bibinfo{title}{From organized {High-Throughput} data to
		phenomenological theory using machine learning: The example of dielectric
		breakdown}.
	\newblock \emph{\bibinfo{journal}{Chem. Mater.}} \textbf{\bibinfo{volume}{28}},
	\bibinfo{pages}{1304--1311} (\bibinfo{year}{2016}).
	
	\bibitem{21_Kim2016-fc}
	\bibinfo{author}{Kim, C.}, \bibinfo{author}{Pilania, G.} \&
	\bibinfo{author}{Ramprasad, R.}
	\newblock \bibinfo{title}{Machine learning assisted predictions of intrinsic
		dielectric breakdown strength of {ABX3} perovskites}.
	\newblock \emph{\bibinfo{journal}{J. Phys. Chem. C}}
	\textbf{\bibinfo{volume}{120}}, \bibinfo{pages}{14575--14580}
	(\bibinfo{year}{2016}).
	
	\bibitem{16_Schmidt2009-pn}
	\bibinfo{author}{Schmidt, M.} \& \bibinfo{author}{Lipson, H.}
	\newblock \bibinfo{title}{Distilling free-form natural laws from experimental
		data}.
	\newblock \emph{\bibinfo{journal}{Science}} \textbf{\bibinfo{volume}{324}},
	\bibinfo{pages}{81--85} (\bibinfo{year}{2009}).
	
	\bibitem{17_Ghiringhelli2015-bb}
	\bibinfo{author}{Ghiringhelli, L.~M.}, \bibinfo{author}{Vybiral, J.},
	\bibinfo{author}{Levchenko, S.~V.}, \bibinfo{author}{Draxl, C.} \&
	\bibinfo{author}{Scheffler, M.}
	\newblock \bibinfo{title}{Big data of materials science: critical role of the
		descriptor}.
	\newblock \emph{\bibinfo{journal}{Phys. Rev. Lett.}}
	\textbf{\bibinfo{volume}{114}}, \bibinfo{pages}{105503}
	(\bibinfo{year}{2015}).
	
	\bibitem{18_Ghiringhelli2017-wt}
	\bibinfo{author}{Ghiringhelli, L.~M.} \emph{et~al.}
	\newblock \bibinfo{title}{Learning physical descriptors for materials science
		by compressed sensing}.
	\newblock \emph{\bibinfo{journal}{New J. Phys.}} \textbf{\bibinfo{volume}{19}},
	\bibinfo{pages}{023017} (\bibinfo{year}{2017}).
	
	\bibitem{19_Lookman2015-xa}
	\bibinfo{author}{Lookman, T.}, \bibinfo{author}{Alexander, F.~J.} \&
	\bibinfo{author}{Rajan, K.}
	\newblock \emph{\bibinfo{title}{Information Science for Materials Discovery and
			Design}} (\bibinfo{publisher}{Springer}, \bibinfo{year}{2015}).
	
	\bibitem{22_Goldsmith2017-vg}
	\bibinfo{author}{Goldsmith, B.~R.} \emph{et~al.}
	\newblock \bibinfo{title}{Uncovering structure-property relationships of
		materials by subgroup discovery}.
	\newblock \emph{\bibinfo{journal}{New J. Phys.}} \textbf{\bibinfo{volume}{19}},
	\bibinfo{pages}{013031} (\bibinfo{year}{2017}).
	
	\bibitem{23_Bialon2016-gf}
	\bibinfo{author}{Bialon, A.~F.}, \bibinfo{author}{Hammerschmidt, T.} \&
	\bibinfo{author}{Drautz, R.}
	\newblock \bibinfo{title}{{Three-Parameter} {Crystal-Structure} prediction for
		sp-d-valent compounds}.
	\newblock \emph{\bibinfo{journal}{Chem. Mater.}} \textbf{\bibinfo{volume}{28}},
	\bibinfo{pages}{2550--2556} (\bibinfo{year}{2016}).
	
	\bibitem{24_noauthor_2008-qb}
	\bibinfo{title}{Pearson's crystal data: Crystal structure database for
		inorganic compounds}.
	\newblock \emph{\bibinfo{journal}{Choice Reviews Online}}
	\textbf{\bibinfo{volume}{45}}, \bibinfo{pages}{45--3800--45--3800}
	(\bibinfo{year}{2008}).
	
	\bibitem{25_Oliynyk2016-bd}
	\bibinfo{author}{Oliynyk, A.~O.} \emph{et~al.}
	\newblock \bibinfo{title}{{High-Throughput} {Machine-Learning-Driven} synthesis
		of {Full-Heusler} compounds}.
	\newblock \emph{\bibinfo{journal}{Chem. Mater.}} \textbf{\bibinfo{volume}{28}},
	\bibinfo{pages}{7324--7331} (\bibinfo{year}{2016}).
	
	\bibitem{26_noauthor_undated-rh}
	\bibinfo{title}{{ASM} international the materials information society - {ASM}
		international}.
	\newblock \bibinfo{howpublished}{\url{http://www.asminternational.org/}}.
	\newblock \bibinfo{note}{Accessed: 2017-6-23}.
	
	\bibitem{27_Dey2014-sq}
	\bibinfo{author}{Dey, P.} \emph{et~al.}
	\newblock \bibinfo{title}{Informatics-aided bandgap engineering for solar
		materials}.
	\newblock \emph{\bibinfo{journal}{Comput. Mater. Sci.}}
	\textbf{\bibinfo{volume}{83}}, \bibinfo{pages}{185--195}
	(\bibinfo{year}{2014}).
	
	\bibitem{28_Ward2016-ia}
	\bibinfo{author}{Ward, L.}, \bibinfo{author}{Agrawal, A.},
	\bibinfo{author}{Choudhary, A.} \& \bibinfo{author}{Wolverton, C.}
	\newblock \bibinfo{title}{A general-purpose machine learning framework for
		predicting properties of inorganic materials}.
	\newblock \emph{\bibinfo{journal}{npj Comput. Mater.}}
	\textbf{\bibinfo{volume}{2}}, \bibinfo{pages}{npjcompumats201628}
	(\bibinfo{year}{2016}).
	
	\bibitem{29_Lee2016-lr}
	\bibinfo{author}{Lee, J.}, \bibinfo{author}{Seko, A.},
	\bibinfo{author}{Shitara, K.}, \bibinfo{author}{Nakayama, K.} \&
	\bibinfo{author}{Tanaka, I.}
	\newblock \bibinfo{title}{Prediction model of band gap for inorganic compounds
		by combination of density functional theory calculations and machine learning
		techniques}.
	\newblock \emph{\bibinfo{journal}{Phys. Rev. B Condens. Matter}}
	\textbf{\bibinfo{volume}{93}}, \bibinfo{pages}{115104}
	(\bibinfo{year}{2016}).
	
	\bibitem{30_Pilania2016-ig}
	\bibinfo{author}{Pilania, G.} \emph{et~al.}
	\newblock \bibinfo{title}{Machine learning bandgaps of double perovskites}.
	\newblock \emph{\bibinfo{journal}{Sci. Rep.}} \textbf{\bibinfo{volume}{6}},
	\bibinfo{pages}{19375} (\bibinfo{year}{2016}).
	
	\bibitem{31_Pilania2017-kq}
	\bibinfo{author}{Pilania, G.}, \bibinfo{author}{Gubernatis, J.~E.} \&
	\bibinfo{author}{Lookman, T.}
	\newblock \bibinfo{title}{Multi-fidelity machine learning models for accurate
		bandgap predictions of solids}.
	\newblock \emph{\bibinfo{journal}{Comput. Mater. Sci.}}
	\textbf{\bibinfo{volume}{129}}, \bibinfo{pages}{156--163}
	(\bibinfo{year}{2017}).
	
	\bibitem{32_Faber2016-ab}
	\bibinfo{author}{Faber, F.~A.}, \bibinfo{author}{Lindmaa, A.},
	\bibinfo{author}{von Lilienfeld, O.~A.} \& \bibinfo{author}{Armiento, R.}
	\newblock \bibinfo{title}{Machine learning energies of 2 million elpasolite
		({ABC$_{2}$D$_{6}$}) crystals}.
	\newblock \emph{\bibinfo{journal}{Phys. Rev. Lett.}}
	\textbf{\bibinfo{volume}{117}}, \bibinfo{pages}{135502}
	(\bibinfo{year}{2016}).
	
	\bibitem{33_Meredig2014-vi}
	\bibinfo{author}{Meredig, B.} \emph{et~al.}
	\newblock \bibinfo{title}{Combinatorial screening for new materials in
		unconstrained composition space with machine learning}.
	\newblock \emph{\bibinfo{journal}{Phys. Rev. B Condens. Matter}}
	\textbf{\bibinfo{volume}{89}}, \bibinfo{pages}{094104}
	(\bibinfo{year}{2014}).
	
	\bibitem{34_Deml2016-dt}
	\bibinfo{author}{Deml, A.~M.}, \bibinfo{author}{{O'}Hayre, R.},
	\bibinfo{author}{Wolverton, C.} \& \bibinfo{author}{Stevanovi{\'c}, V.}
	\newblock \bibinfo{title}{Predicting density functional theory total energies
		and enthalpies of formation of metal-nonmetal compounds by linear
		regression}.
	\newblock \emph{\bibinfo{journal}{Phys. Rev. B Condens. Matter}}
	\textbf{\bibinfo{volume}{93}}, \bibinfo{pages}{085142}
	(\bibinfo{year}{2016}).
	
	\bibitem{35_Legrain2017-kp}
	\bibinfo{author}{Legrain, F.}, \bibinfo{author}{Carrete, J.},
	\bibinfo{author}{van Roekeghem, A.}, \bibinfo{author}{Curtarolo, S.} \&
	\bibinfo{author}{Mingo, N.}
	\newblock \bibinfo{title}{How the chemical composition alone can predict
		vibrational free energies and entropies of solids}.
	\newblock \emph{\bibinfo{journal}{Chem. Mater.}}  (\bibinfo{year}{2017}).
	
	\bibitem{36_Medasani2016-ss}
	\bibinfo{author}{Medasani, B.} \emph{et~al.}
	\newblock \bibinfo{title}{Predicting defect behavior in {B2} intermetallics by
		merging ab initio modeling and machine learning}.
	\newblock \emph{\bibinfo{journal}{npj Comput. Mater.}}
	\textbf{\bibinfo{volume}{2}}, \bibinfo{pages}{1} (\bibinfo{year}{2016}).
	
	\bibitem{37_Seko2014-ov}
	\bibinfo{author}{Seko, A.}, \bibinfo{author}{Maekawa, T.},
	\bibinfo{author}{Tsuda, K.} \& \bibinfo{author}{Tanaka, I.}
	\newblock \bibinfo{title}{Machine learning with systematic density-functional
		theory calculations: Application to melting temperatures of single- and
		binary-component solids}.
	\newblock \emph{\bibinfo{journal}{Phys. Rev. B Condens. Matter}}
	\textbf{\bibinfo{volume}{89}}, \bibinfo{pages}{054303}
	(\bibinfo{year}{2014}).
	
	\bibitem{38_Pilania2015-yw}
	\bibinfo{author}{Pilania, G.}, \bibinfo{author}{Gubernatis, J.~E.} \&
	\bibinfo{author}{Lookman, T.}
	\newblock \bibinfo{title}{Structure classification and melting temperature
		prediction in octet {AB} solids via machine learning}.
	\newblock \emph{\bibinfo{journal}{Phys. Rev. B Condens. Matter}}
	\textbf{\bibinfo{volume}{91}}, \bibinfo{pages}{214302}
	(\bibinfo{year}{2015}).
	
	\bibitem{39_Chatterjee2007-yz}
	\bibinfo{author}{Chatterjee, S.}, \bibinfo{author}{Murugananth, M.} \&
	\bibinfo{author}{Bhadeshia, H. K. D.~H.}
	\newblock \bibinfo{title}{$\delta$ {TRIP} steel}.
	\newblock \emph{\bibinfo{journal}{Mater. Sci. Technol.}}
	\textbf{\bibinfo{volume}{23}}, \bibinfo{pages}{819--827}
	(\bibinfo{year}{2007}).
	
	\bibitem{40_De_Jong2016-ki}
	\bibinfo{author}{De~Jong, M.} \emph{et~al.}
	\newblock \bibinfo{title}{A statistical learning framework for materials
		science: Application to elastic moduli of k-nary inorganic polycrystalline
		compounds}.
	\newblock \emph{\bibinfo{journal}{Sci. Rep.}} \textbf{\bibinfo{volume}{6}},
	\bibinfo{pages}{34256} (\bibinfo{year}{2016}).
	
	\bibitem{41_Aryal2014-ab}
	\bibinfo{author}{Aryal, S.}, \bibinfo{author}{Sakidja, R.},
	\bibinfo{author}{Barsoum, M.~W.} \& \bibinfo{author}{Ching, W.-Y.}
	\newblock \bibinfo{title}{A genomic approach to the stability, elastic, and
		electronic properties of the {MAX} phases}.
	\newblock \emph{\bibinfo{journal}{Phys. Status Solidi}}
	\textbf{\bibinfo{volume}{251}}, \bibinfo{pages}{1480--1497}
	(\bibinfo{year}{2014}).
	
	\bibitem{42_Seko2015-qe}
	\bibinfo{author}{Seko, A.} \emph{et~al.}
	\newblock \bibinfo{title}{Prediction of {Low-Thermal-Conductivity} compounds
		with {First-Principles} anharmonic {Lattice-Dynamics} calculations and
		bayesian optimization}.
	\newblock \emph{\bibinfo{journal}{Phys. Rev. Lett.}}
	\textbf{\bibinfo{volume}{115}}, \bibinfo{pages}{205901}
	(\bibinfo{year}{2015}).
	
	\bibitem{43_Li2017-oq}
	\bibinfo{author}{Li, Z.}, \bibinfo{author}{Ma, X.} \& \bibinfo{author}{Xin, H.}
	\newblock \bibinfo{title}{Feature engineering of machine-learning chemisorption
		models for catalyst design}.
	\newblock \emph{\bibinfo{journal}{Catal. Today}} \textbf{\bibinfo{volume}{280,
			Part 2}}, \bibinfo{pages}{232--238} (\bibinfo{year}{2017}).
	
	\bibitem{44_Hong2016-lt}
	\bibinfo{author}{Hong, W.~T.}, \bibinfo{author}{Welsch, R.~E.} \&
	\bibinfo{author}{Shao-Horn, Y.}
	\newblock \bibinfo{title}{Descriptors of {Oxygen-Evolution} activity for
		oxides: A statistical evaluation}.
	\newblock \emph{\bibinfo{journal}{J. Phys. Chem. C}}
	\textbf{\bibinfo{volume}{120}}, \bibinfo{pages}{78--86}
	(\bibinfo{year}{2016}).
	
	\bibitem{45_Pilania2017-ci}
	\bibinfo{author}{Pilania, G.} \emph{et~al.}
	\newblock \bibinfo{title}{Using machine learning to identify factors that
		govern amorphization of irradiated pyrochlores}.
	\newblock \emph{\bibinfo{journal}{Chem. Mater.}} \textbf{\bibinfo{volume}{29}},
	\bibinfo{pages}{2574--2583} (\bibinfo{year}{2017}).
	
	\bibitem{46_Xue2016-us}
	\bibinfo{author}{Xue, D.} \emph{et~al.}
	\newblock \bibinfo{title}{Accelerated search for materials with targeted
		properties by adaptive design}.
	\newblock \emph{\bibinfo{journal}{Nat. Commun.}} \textbf{\bibinfo{volume}{7}},
	\bibinfo{pages}{11241} (\bibinfo{year}{2016}).
	
	\bibitem{47_Xue2016-gu}
	\bibinfo{author}{Xue, D.} \emph{et~al.}
	\newblock \bibinfo{title}{Accelerated search for {BaTiO3-based} piezoelectrics
		with vertical morphotropic phase boundary using bayesian learning}.
	\newblock \emph{\bibinfo{journal}{Proc. Natl. Acad. Sci. U. S. A.}}
	\textbf{\bibinfo{volume}{113}}, \bibinfo{pages}{13301--13306}
	(\bibinfo{year}{2016}).
	
	\bibitem{48_Ashton2016-fe}
	\bibinfo{author}{Ashton, M.}, \bibinfo{author}{Hennig, R.~G.},
	\bibinfo{author}{Broderick, S.~R.}, \bibinfo{author}{Rajan, K.} \&
	\bibinfo{author}{Sinnott, S.~B.}
	\newblock \bibinfo{title}{Computational discovery of stable {M$_2$AX} phases}.
	\newblock \emph{\bibinfo{journal}{Phys. Rev. B Condens. Matter}}
	\textbf{\bibinfo{volume}{94}}, \bibinfo{pages}{20} (\bibinfo{year}{2016}).
	
	\bibitem{49_Pilania2016-at}
	\bibinfo{author}{Pilania, G.}, \bibinfo{author}{Balachandran, P.~V.},
	\bibinfo{author}{Kim, C.} \& \bibinfo{author}{Lookman, T.}
	\newblock \bibinfo{title}{Finding new perovskite halides via machine learning}.
	\newblock \emph{\bibinfo{journal}{Frontiers in Materials}}
	\textbf{\bibinfo{volume}{3}}, \bibinfo{pages}{19} (\bibinfo{year}{2016}).
	
	\bibitem{50_Fernandez2014-bq}
	\bibinfo{author}{Fernandez, M.}, \bibinfo{author}{Boyd, P.~G.},
	\bibinfo{author}{Daff, T.~D.}, \bibinfo{author}{Aghaji, M.~Z.} \&
	\bibinfo{author}{Woo, T.~K.}
	\newblock \bibinfo{title}{Rapid and accurate machine learning recognition of
		high performing metal organic frameworks for {CO2} capture}.
	\newblock \emph{\bibinfo{journal}{J. Phys. Chem. Lett.}}
	\textbf{\bibinfo{volume}{5}}, \bibinfo{pages}{3056--3060}
	(\bibinfo{year}{2014}).
	
	\bibitem{51_Emery2016-cx}
	\bibinfo{author}{Emery, A.~A.}, \bibinfo{author}{Saal, J.~E.},
	\bibinfo{author}{Kirklin, S.}, \bibinfo{author}{Hegde, V.~I.} \&
	\bibinfo{author}{Wolverton, C.}
	\newblock \bibinfo{title}{{High-Throughput} computational screening of
		perovskites for thermochemical water splitting applications}.
	\newblock \emph{\bibinfo{journal}{Chem. Mater.}} \textbf{\bibinfo{volume}{28}},
	\bibinfo{pages}{5621--5634} (\bibinfo{year}{2016}).
	
	\bibitem{52_Kalidindi2016-ss}
	\bibinfo{author}{Kalidindi, S.~R.} \emph{et~al.}
	\newblock \bibinfo{title}{Role of materials data science and informatics in
		accelerated materials innovation}.
	\newblock \emph{\bibinfo{journal}{MRS Bull.}} \textbf{\bibinfo{volume}{41}},
	\bibinfo{pages}{596--602} (\bibinfo{year}{2016}).
	
	\bibitem{53_Brough2017-zp}
	\bibinfo{author}{Brough, D.~B.}, \bibinfo{author}{Kannan, A.},
	\bibinfo{author}{Haaland, B.}, \bibinfo{author}{Bucknall, D.~G.} \&
	\bibinfo{author}{Kalidindi, S.~R.}
	\newblock \bibinfo{title}{Extraction of {Process-Structure} evolution linkages
		from x-ray scattering measurements using dimensionality reduction and time
		series analysis}.
	\newblock \emph{\bibinfo{journal}{Integr Mater Manuf Innov}}
	\textbf{\bibinfo{volume}{6}}, \bibinfo{pages}{147--159}
	(\bibinfo{year}{2017}).
	
	\bibitem{54_Kalidindi2015-tk}
	\bibinfo{author}{Kalidindi, S.~R.}, \bibinfo{author}{Gomberg, J.~A.},
	\bibinfo{author}{Trautt, Z.~T.} \& \bibinfo{author}{Becker, C.~A.}
	\newblock \bibinfo{title}{Application of data science tools to quantify and
		distinguish between structures and models in molecular dynamics datasets}.
	\newblock \emph{\bibinfo{journal}{Nanotechnology}}
	\textbf{\bibinfo{volume}{26}}, \bibinfo{pages}{344006}
	(\bibinfo{year}{2015}).
	
	\bibitem{55_Gupta2015-fs}
	\bibinfo{author}{Gupta, A.}, \bibinfo{author}{Cecen, A.},
	\bibinfo{author}{Goyal, S.}, \bibinfo{author}{Singh, A.~K.} \&
	\bibinfo{author}{Kalidindi, S.~R.}
	\newblock \bibinfo{title}{Structure--property linkages using a data science
		approach: Application to a non-metallic inclusion/steel composite system}.
	\newblock \emph{\bibinfo{journal}{Acta Mater.}} \textbf{\bibinfo{volume}{91}},
	\bibinfo{pages}{239--254} (\bibinfo{year}{2015}).
	
	\bibitem{56_Brough2017-gh}
	\bibinfo{author}{Brough, D.~B.}, \bibinfo{author}{Wheeler, D.},
	\bibinfo{author}{Warren, J.~A.} \& \bibinfo{author}{Kalidindi, S.~R.}
	\newblock \bibinfo{title}{Microstructure-based knowledge systems for capturing
		process-structure evolution linkages}.
	\newblock \emph{\bibinfo{journal}{Curr. Opin. Solid State Mater. Sci.}}
	\textbf{\bibinfo{volume}{21}}, \bibinfo{pages}{129--140}
	(\bibinfo{year}{2017}).
	
	\bibitem{57_Panchal2013-bf}
	\bibinfo{author}{Panchal, J.~H.}, \bibinfo{author}{Kalidindi, S.~R.} \&
	\bibinfo{author}{McDowell, D.~L.}
	\newblock \bibinfo{title}{Key computational modeling issues in integrated
		computational materials engineering}.
	\newblock \emph{\bibinfo{journal}{Comput. Aided Des. Appl.}}
	\textbf{\bibinfo{volume}{45}}, \bibinfo{pages}{4--25} (\bibinfo{year}{2013}).
	
	\bibitem{58_Brough2017-vy}
	\bibinfo{author}{Brough, D.~B.}, \bibinfo{author}{Wheeler, D.} \&
	\bibinfo{author}{Kalidindi, S.~R.}
	\newblock \bibinfo{title}{Materials knowledge systems in python---a data
		science framework for accelerated development of hierarchical materials}.
	\newblock \emph{\bibinfo{journal}{Integr Mater Manuf Innov}}
	\textbf{\bibinfo{volume}{6}}, \bibinfo{pages}{36--53} (\bibinfo{year}{2017}).
	
	\bibitem{59_Kalidindi2012-xw}
	\bibinfo{author}{Kalidindi, S.~R.}
	\newblock \bibinfo{title}{Computationally efficient, fully coupled multiscale
		modeling of materials phenomena using calibrated localization linkages}.
	\newblock \emph{\bibinfo{journal}{International Scholarly Research Notices}}
	\textbf{\bibinfo{volume}{2012}} (\bibinfo{year}{2012}).
	
	\bibitem{60_Adamson1974-uy}
	\bibinfo{author}{Adamson, G.~W.} \& \bibinfo{author}{Bush, J.~A.}
	\newblock \bibinfo{title}{Method for relating the structure and properties of
		chemical compounds}.
	\newblock \emph{\bibinfo{journal}{Nature}} \textbf{\bibinfo{volume}{248}},
	\bibinfo{pages}{406--407} (\bibinfo{year}{1974}).
	
	\bibitem{61_Adamson1974-jp}
	\bibinfo{author}{Adamson, G.~W.}, \bibinfo{author}{Bush, J.~A.},
	\bibinfo{author}{McLure, A. H.~W.} \& \bibinfo{author}{Lynch, M.~F.}
	\newblock \bibinfo{title}{An evaluation of a substructure search screen system
		based on {Bond-Centered} fragments}.
	\newblock \emph{\bibinfo{journal}{J. Chem. Doc.}}
	\textbf{\bibinfo{volume}{14}}, \bibinfo{pages}{44--48}
	(\bibinfo{year}{1974}).
	
	\bibitem{62_Judson2009-zq}
	\bibinfo{author}{Judson, P.}
	\newblock \emph{\bibinfo{title}{{Knowledge-Based} Expert Systems in Chemistry:
			Not Counting on Computers}} (\bibinfo{publisher}{Royal Society of Chemistry},
	\bibinfo{year}{2009}).
	
	\bibitem{68_Mannodi-Kanakkithodi2016-vc}
	\bibinfo{author}{Mannodi-Kanakkithodi, A.}, \bibinfo{author}{Pilania, G.},
	\bibinfo{author}{Huan, T.~D.}, \bibinfo{author}{Lookman, T.} \&
	\bibinfo{author}{Ramprasad, R.}
	\newblock \bibinfo{title}{Machine learning strategy for accelerated design of
		polymer dielectrics}.
	\newblock \emph{\bibinfo{journal}{Sci. Rep.}} \textbf{\bibinfo{volume}{6}},
	\bibinfo{pages}{20952} (\bibinfo{year}{2016}).
	
	\bibitem{63_Huan2016-kk}
	\bibinfo{author}{Huan, T.~D.} \emph{et~al.}
	\newblock \bibinfo{title}{A polymer dataset for accelerated property prediction
		and design}.
	\newblock \emph{\bibinfo{journal}{Sci Data}} \textbf{\bibinfo{volume}{3}},
	\bibinfo{pages}{160012} (\bibinfo{year}{2016}).
	
	\bibitem{64_Mannodi-Kanakkithodi2016-zv}
	\bibinfo{author}{Mannodi-Kanakkithodi, A.} \emph{et~al.}
	\newblock \bibinfo{title}{Rational {Co-Design} of polymer dielectrics for
		energy storage}.
	\newblock \emph{\bibinfo{journal}{Adv. Mater.}} \textbf{\bibinfo{volume}{28}},
	\bibinfo{pages}{6277--6291} (\bibinfo{year}{2016}).
	
	\bibitem{65_Treich2017-xn}
	\bibinfo{author}{Treich, G.~M.} \emph{et~al.}
	\newblock \bibinfo{title}{A rational co-design approach to the creation of new
		dielectric polymers with high energy density}.
	\newblock \emph{\bibinfo{journal}{IEEE Trans. Dielectr. Electr. Insul.}}
	\textbf{\bibinfo{volume}{24}}, \bibinfo{pages}{732--743}
	(\bibinfo{year}{2017}).
	
	\bibitem{66_Huan2016-kl}
	\bibinfo{author}{Huan, T.~D.} \emph{et~al.}
	\newblock \bibinfo{title}{Advanced polymeric dielectrics for high energy
		density applications}.
	\newblock \emph{\bibinfo{journal}{Prog. Mater Sci.}}
	\textbf{\bibinfo{volume}{83}}, \bibinfo{pages}{236--269}
	(\bibinfo{year}{2016}).
	
	\bibitem{67_Sharma2014-re}
	\bibinfo{author}{Sharma, V.} \emph{et~al.}
	\newblock \bibinfo{title}{Rational design of all organic polymer dielectrics}.
	\newblock \emph{\bibinfo{journal}{Nat. Commun.}} \textbf{\bibinfo{volume}{5}},
	\bibinfo{pages}{4845} (\bibinfo{year}{2014}).
	
	\bibitem{114_Lorenzini}
	\bibinfo{author}{Lorenzini, R.~G.}, \bibinfo{author}{Kline, W.~M.},
	\bibinfo{author}{Wang, C.~C.}, \bibinfo{author}{R, R.} \& \bibinfo{author}{A,
		S.~G.}
	\newblock \bibinfo{title}{The rational design of polyurea \& polyurethane
		dielectric materials}.
	\newblock \emph{\bibinfo{journal}{Polymer}} \textbf{\bibinfo{volume}{54}},
	\bibinfo{pages}{3529} (\bibinfo{year}{2013}).
	
	\bibitem{115_Chunsheng}
	\bibinfo{author}{Liu, C.~S.}, \bibinfo{author}{G, P.}, \bibinfo{author}{C, W.}
	\& \bibinfo{author}{R, R.}
	\newblock \bibinfo{title}{How critical are the van der waals interactions in
		polymer crystals?}
	\newblock \emph{\bibinfo{journal}{The Journal of Physical Chemistry A}}
	\textbf{\bibinfo{volume}{116}}, \bibinfo{pages}{9347} (\bibinfo{year}{2012}).
	
	\bibitem{69_Pilania2013-ec}
	\bibinfo{author}{Pilania, G.}, \bibinfo{author}{Wang, C.},
	\bibinfo{author}{Jiang, X.}, \bibinfo{author}{Rajasekaran, S.} \&
	\bibinfo{author}{Ramprasad, R.}
	\newblock \bibinfo{title}{Accelerating materials property predictions using
		machine learning}.
	\newblock \emph{\bibinfo{journal}{Sci. Rep.}} \textbf{\bibinfo{volume}{3}},
	\bibinfo{pages}{2810} (\bibinfo{year}{2013}).
	
	\bibitem{70_T_D_Huan_A_Mannodi-Kanakkithodi_R_Ramprasad2015-rg}
	\bibinfo{author}{{T. D. Huan, A. Mannodi-Kanakkithodi, R. Ramprasad}}.
	\newblock \bibinfo{title}{Accelerated materials property predictions and design
		using motif-based fingerprints}.
	\newblock \emph{\bibinfo{journal}{Phys. Rev. B Condens. Matter}}
	\textbf{\bibinfo{volume}{92}} (\bibinfo{year}{2015}).
	
	\bibitem{71_A_Mannodi-Kanakkithodi_T_D_Huan_R_Ramprasad_undated-xm}
	\bibinfo{author}{{A Mannodi-Kanakkithodi, T D Huan, R. Ramprasad}}.
	\newblock \bibinfo{title}{Mining materials design rules from data: The example
		of polymer dielectrics}.
	\newblock \emph{\bibinfo{journal}{(Under Review)}} .
	
	\bibitem{72_noauthor_undated-cb}
	\bibinfo{title}{{PolymerGenome}}.
	\newblock \bibinfo{howpublished}{\url{http://polymergenome.org}}.
	
	\bibitem{73_Hautier2010-ou}
	\bibinfo{author}{Hautier, G.}, \bibinfo{author}{Fischer, C.~C.},
	\bibinfo{author}{Jain, A.}, \bibinfo{author}{Mueller, T.} \&
	\bibinfo{author}{Ceder, G.}
	\newblock \bibinfo{title}{Finding nature's missing ternary oxide compounds
		using machine learning and density functional theory}.
	\newblock \emph{\bibinfo{journal}{Chem. Mater.}} \textbf{\bibinfo{volume}{22}},
	\bibinfo{pages}{3762--3767} (\bibinfo{year}{2010}).
	
	\bibitem{74_Behler2007-ig}
	\bibinfo{author}{Behler, J.} \& \bibinfo{author}{Parrinello, M.}
	\newblock \bibinfo{title}{Generalized neural-network representation of
		high-dimensional potential-energy surfaces}.
	\newblock \emph{\bibinfo{journal}{Phys. Rev. Lett.}}
	\textbf{\bibinfo{volume}{98}}, \bibinfo{pages}{146401}
	(\bibinfo{year}{2007}).
	
	\bibitem{75_Behler2008-tn}
	\bibinfo{author}{Behler, J.}, \bibinfo{author}{Marton{\'a}k, R.},
	\bibinfo{author}{Donadio, D.} \& \bibinfo{author}{Parrinello, M.}
	\newblock \bibinfo{title}{Metadynamics simulations of the high-pressure phases
		of silicon employing a high-dimensional neural network potential}.
	\newblock \emph{\bibinfo{journal}{Phys. Rev. Lett.}}
	\textbf{\bibinfo{volume}{100}}, \bibinfo{pages}{185501}
	(\bibinfo{year}{2008}).
	
	\bibitem{76_Behler2014-zf}
	\bibinfo{author}{Behler, J.}
	\newblock \bibinfo{title}{Representing potential energy surfaces by
		high-dimensional neural network potentials}.
	\newblock \emph{\bibinfo{journal}{J. Phys. Condens. Matter}}
	\textbf{\bibinfo{volume}{26}}, \bibinfo{pages}{183001}
	(\bibinfo{year}{2014}).
	
	\bibitem{77_Bartok2010-zz}
	\bibinfo{author}{Bart{\'o}k, A.~P.}, \bibinfo{author}{Payne, M.~C.},
	\bibinfo{author}{Kondor, R.} \& \bibinfo{author}{Cs{\'a}nyi, G.}
	\newblock \bibinfo{title}{Gaussian approximation potentials: the accuracy of
		quantum mechanics, without the electrons}.
	\newblock \emph{\bibinfo{journal}{Phys. Rev. Lett.}}
	\textbf{\bibinfo{volume}{104}}, \bibinfo{pages}{136403}
	(\bibinfo{year}{2010}).
	
	\bibitem{78_Rupp2012-in}
	\bibinfo{author}{Rupp, M.}, \bibinfo{author}{Tkatchenko, A.},
	\bibinfo{author}{M{\"u}ller, K.-R.} \& \bibinfo{author}{von Lilienfeld,
		O.~A.}
	\newblock \bibinfo{title}{Fast and accurate modeling of molecular atomization
		energies with machine learning}.
	\newblock \emph{\bibinfo{journal}{Phys. Rev. Lett.}}
	\textbf{\bibinfo{volume}{108}}, \bibinfo{pages}{058301}
	(\bibinfo{year}{2012}).
	
	\bibitem{79_Chmiela2017-pp}
	\bibinfo{author}{Chmiela, S.} \emph{et~al.}
	\newblock \bibinfo{title}{Machine learning of accurate energy-conserving
		molecular force fields}.
	\newblock \emph{\bibinfo{journal}{Sci Adv}} \textbf{\bibinfo{volume}{3}},
	\bibinfo{pages}{e1603015} (\bibinfo{year}{2017}).
	
	\bibitem{80_Bartok2013-tj}
	\bibinfo{author}{Bart{\'o}k, A.~P.}, \bibinfo{author}{Kondor, R.} \&
	\bibinfo{author}{Cs{\'a}nyi, G.}
	\newblock \bibinfo{title}{On representing chemical environments}.
	\newblock \emph{\bibinfo{journal}{Phys. Rev. B Condens. Matter}}
	\textbf{\bibinfo{volume}{87}}, \bibinfo{pages}{184115}
	(\bibinfo{year}{2013}).
	
	\bibitem{81_Szlachta2014-dp}
	\bibinfo{author}{Szlachta, W.~J.}, \bibinfo{author}{Bart{\'o}k, A.~P.} \&
	\bibinfo{author}{Cs{\'a}nyi, G.}
	\newblock \bibinfo{title}{Accuracy and transferability of gaussian
		approximation potential models for tungsten}.
	\newblock \emph{\bibinfo{journal}{Phys. Rev. B Condens. Matter}}
	\textbf{\bibinfo{volume}{90}}, \bibinfo{pages}{104108}
	(\bibinfo{year}{2014}).
	
	\bibitem{82_Bartok2015-ap}
	\bibinfo{author}{Bart{\'o}k, A.~P.} \& \bibinfo{author}{Cs{\'a}nyi, G.}
	\newblock \bibinfo{title}{Gaussian approximation potentials: A brief tutorial
		introduction}.
	\newblock \emph{\bibinfo{journal}{Int. J. Quantum Chem.}}
	\textbf{\bibinfo{volume}{115}}, \bibinfo{pages}{1051--1057}
	(\bibinfo{year}{2015}).
	
	\bibitem{83_Deringer2017-sq}
	\bibinfo{author}{Deringer, V.~L.} \& \bibinfo{author}{Cs{\'a}nyi, G.}
	\newblock \bibinfo{title}{Machine learning based interatomic potential for
		amorphous carbon}.
	\newblock \emph{\bibinfo{journal}{Phys. Rev. B Condens. Matter}}
	\textbf{\bibinfo{volume}{95}}, \bibinfo{pages}{094203}
	(\bibinfo{year}{2017}).
	
	\bibitem{84_Jindal2017-ov}
	\bibinfo{author}{Jindal, S.}, \bibinfo{author}{Chiriki, S.} \&
	\bibinfo{author}{Bulusu, S.~S.}
	\newblock \bibinfo{title}{Spherical harmonics based descriptor for neural
		network potentials: Structure and dynamics of {Au$_{147}$} nanocluster}.
	\newblock \emph{\bibinfo{journal}{J. Chem. Phys.}}
	\textbf{\bibinfo{volume}{146}}, \bibinfo{pages}{204301}
	(\bibinfo{year}{2017}).
	
	\bibitem{85_Thompson_undated-ka}
	\bibinfo{author}{Thompson, A.}, \bibinfo{author}{Swiler, L.},
	\bibinfo{author}{Trott, C.}, \bibinfo{author}{Foiles, S.} \&
	\bibinfo{author}{Tucker, G.}
	\newblock \bibinfo{title}{Spectral neighbor analysis method for automated
		generation of quantum-accurate interatomic potentials}.
	\newblock \emph{\bibinfo{journal}{J. Comput. Phys.}}
	\textbf{\bibinfo{volume}{285}}, \bibinfo{pages}{316 -- 330}
	(\bibinfo{year}{2015}).
	
	\bibitem{86_Rupp2015-si}
	\bibinfo{author}{Rupp, M.}
	\newblock \bibinfo{title}{Machine learning for quantum mechanics in a
		nutshell}.
	\newblock \emph{\bibinfo{journal}{Int. J. Quantum Chem.}}
	\textbf{\bibinfo{volume}{115}}, \bibinfo{pages}{1058--1073}
	(\bibinfo{year}{2015}).
	
	\bibitem{87_Li2015-nv}
	\bibinfo{author}{Li, Z.}, \bibinfo{author}{Kermode, J.~R.} \&
	\bibinfo{author}{De~Vita, A.}
	\newblock \bibinfo{title}{Molecular dynamics with on-the-fly machine learning
		of quantum-mechanical forces}.
	\newblock \emph{\bibinfo{journal}{Phys. Rev. Lett.}}
	\textbf{\bibinfo{volume}{114}}, \bibinfo{pages}{096405}
	(\bibinfo{year}{2015}).
	
	\bibitem{88_Botu2015-jb}
	\bibinfo{author}{Botu, V.} \& \bibinfo{author}{Ramprasad, R.}
	\newblock \bibinfo{title}{Learning scheme to predict atomic forces and
		accelerate materials simulations}.
	\newblock \emph{\bibinfo{journal}{Phys. Rev. B Condens. Matter}}
	\textbf{\bibinfo{volume}{92}}, \bibinfo{pages}{094306}
	(\bibinfo{year}{2015}).
	
	\bibitem{89_Glielmo2017-tr}
	\bibinfo{author}{Glielmo, A.}, \bibinfo{author}{Sollich, P.} \&
	\bibinfo{author}{De~Vita, A.}
	\newblock \bibinfo{title}{Accurate interatomic force fields via machine
		learning with covariant kernels}.
	\newblock \emph{\bibinfo{journal}{Phys. Rev. B Condens. Matter}}
	\textbf{\bibinfo{volume}{95}}, \bibinfo{pages}{214302}
	(\bibinfo{year}{2017}).
	
	\bibitem{90_Botu2015-aa}
	\bibinfo{author}{Botu, V.} \& \bibinfo{author}{Ramprasad, R.}
	\newblock \bibinfo{title}{Adaptive machine learning framework to accelerate ab
		initio molecular dynamics}.
	\newblock \emph{\bibinfo{journal}{Int. J. Quantum Chem.}}
	\textbf{\bibinfo{volume}{115}}, \bibinfo{pages}{1074--1083}
	(\bibinfo{year}{2015}).
	
	\bibitem{91_Botu2017-vk}
	\bibinfo{author}{Botu, V.}, \bibinfo{author}{Chapman, J.} \&
	\bibinfo{author}{Ramprasad, R.}
	\newblock \bibinfo{title}{A study of adatom ripening on an al (111) surface
		with machine learning force fields}.
	\newblock \emph{\bibinfo{journal}{Comput. Mater. Sci.}}
	\textbf{\bibinfo{volume}{129}}, \bibinfo{pages}{332--335}
	(\bibinfo{year}{2017}).
	
	\bibitem{92_Botu2017-kr}
	\bibinfo{author}{Botu, V.}, \bibinfo{author}{Batra, R.},
	\bibinfo{author}{Chapman, J.} \& \bibinfo{author}{Ramprasad, R.}
	\newblock \bibinfo{title}{Machine learning force fields: Construction,
		validation, and outlook}.
	\newblock \emph{\bibinfo{journal}{J. Phys. Chem. C}}
	\textbf{\bibinfo{volume}{121}}, \bibinfo{pages}{511--522}
	(\bibinfo{year}{2017}).
	
	\bibitem{93_Feynman1939-nm}
	\bibinfo{author}{Feynman, R.~P.}
	\newblock \bibinfo{title}{Forces in molecules}.
	\newblock \emph{\bibinfo{journal}{Phys. Rev.}} \textbf{\bibinfo{volume}{56}},
	\bibinfo{pages}{340--343} (\bibinfo{year}{1939}).
	
	\bibitem{94_Bianchini2016-qe}
	\bibinfo{author}{Bianchini, F.}, \bibinfo{author}{Kermode, J.~R.} \&
	\bibinfo{author}{De~Vita, A.}
	\newblock \bibinfo{title}{Modelling defects in {Ni--Al} with {EAM} and {DFT}
		calculations}.
	\newblock \emph{\bibinfo{journal}{Modell. Simul. Mater. Sci. Eng.}}
	\textbf{\bibinfo{volume}{24}}, \bibinfo{pages}{045012}
	(\bibinfo{year}{2016}).
	
	\bibitem{95_De2016-al}
	\bibinfo{author}{De, S.}, \bibinfo{author}{Bart{\'o}k, A.~P.},
	\bibinfo{author}{Cs{\'a}nyi, G.} \& \bibinfo{author}{Ceriotti, M.}
	\newblock \bibinfo{title}{Comparing molecules and solids across structural and
		alchemical space}.
	\newblock \emph{\bibinfo{journal}{Phys. Chem. Chem. Phys.}}
	\textbf{\bibinfo{volume}{18}}, \bibinfo{pages}{13754--13769}
	(\bibinfo{year}{2016}).
	
	\bibitem{96_Snyder2012-aq}
	\bibinfo{author}{Snyder, J.~C.}, \bibinfo{author}{Rupp, M.},
	\bibinfo{author}{Hansen, K.}, \bibinfo{author}{M{\"u}ller, K.-R.} \&
	\bibinfo{author}{Burke, K.}
	\newblock \bibinfo{title}{Finding density functionals with machine learning}.
	\newblock \emph{\bibinfo{journal}{Phys. Rev. Lett.}}
	\textbf{\bibinfo{volume}{108}}, \bibinfo{pages}{253002}
	(\bibinfo{year}{2012}).
	
	\bibitem{97_Snyder2013-bp}
	\bibinfo{author}{Snyder, J.~C.} \emph{et~al.}
	\newblock \bibinfo{title}{Orbital-free bond breaking via machine learning}.
	\newblock \emph{\bibinfo{journal}{J. Chem. Phys.}}
	\textbf{\bibinfo{volume}{139}}, \bibinfo{pages}{224104}
	(\bibinfo{year}{2013}).
	
	\bibitem{98_Snyder2015-ad}
	\bibinfo{author}{Snyder, J.~C.}, \bibinfo{author}{Rupp, M.},
	\bibinfo{author}{M{\"u}ller, K.-R.} \& \bibinfo{author}{Burke, K.}
	\newblock \bibinfo{title}{Nonlinear gradient denoising: Finding accurate
		extrema from inaccurate functional derivatives}.
	\newblock \emph{\bibinfo{journal}{Int. J. Quantum Chem.}}
	\textbf{\bibinfo{volume}{115}}, \bibinfo{pages}{1102--1114}
	(\bibinfo{year}{2015}).
	
	\bibitem{99_Fancher2016-db}
	\bibinfo{author}{Fancher, C.~M.} \emph{et~al.}
	\newblock \bibinfo{title}{Use of bayesian inference in crystallographic
		structure refinement via full diffraction profile analysis}.
	\newblock \emph{\bibinfo{journal}{Sci. Rep.}} \textbf{\bibinfo{volume}{6}},
	\bibinfo{pages}{31625} (\bibinfo{year}{2016}).
	
	\bibitem{100_Kusne2014-dt}
	\bibinfo{author}{Kusne, A.~G.} \emph{et~al.}
	\newblock \bibinfo{title}{On-the-fly machine-learning for high-throughput
		experiments: search for rare-earth-free permanent magnets}.
	\newblock \emph{\bibinfo{journal}{Sci. Rep.}} \textbf{\bibinfo{volume}{4}},
	\bibinfo{pages}{6367} (\bibinfo{year}{2014}).
	
	\bibitem{101_Kusne2015-or}
	\bibinfo{author}{Kusne, A.~G.}, \bibinfo{author}{Keller, D.},
	\bibinfo{author}{Anderson, A.}, \bibinfo{author}{Zaban, A.} \&
	\bibinfo{author}{Takeuchi, I.}
	\newblock \bibinfo{title}{High-throughput determination of structural phase
		diagram and constituent phases using {GRENDEL}}.
	\newblock \emph{\bibinfo{journal}{Nanotechnology}}
	\textbf{\bibinfo{volume}{26}}, \bibinfo{pages}{444002}
	(\bibinfo{year}{2015}).
	
	\bibitem{102_noauthor_2016-al}
	\bibinfo{author}{Hattrick-Simpers, J.~R.}, \bibinfo{author}{Gregoire, J.~M.} \&
	\bibinfo{author}{Kusne, A.~G.}
	\newblock \bibinfo{title}{Perspective: Composition?structure?property
		mapping in high-throughput experiments: Turning data into knowledge}.
	\newblock \emph{\bibinfo{journal}{APL Materials}} \textbf{\bibinfo{volume}{4}},
	\bibinfo{pages}{053211} (\bibinfo{year}{2016}).
	
	\bibitem{103_Bunn2016-wb}
	\bibinfo{author}{Bunn, J.~K.}, \bibinfo{author}{Hu, J.} \&
	\bibinfo{author}{Hattrick-Simpers, J.~R.}
	\newblock \bibinfo{title}{{Semi-Supervised} approach to phase identification
		from combinatorial sample diffraction patterns}.
	\newblock \emph{\bibinfo{journal}{JOM}} \textbf{\bibinfo{volume}{68}},
	\bibinfo{pages}{2116--2125} (\bibinfo{year}{2016}).
	
	\bibitem{104_Lookman2017-fj}
	\bibinfo{author}{Lookman, T.}, \bibinfo{author}{Balachandran, P.~V.},
	\bibinfo{author}{Xue, D.}, \bibinfo{author}{Hogden, J.} \&
	\bibinfo{author}{Theiler, J.}
	\newblock \bibinfo{title}{Statistical inference and adaptive design for
		materials discovery}.
	\newblock \emph{\bibinfo{journal}{Curr. Opin. Solid State Mater. Sci.}}
	\textbf{\bibinfo{volume}{21}}, \bibinfo{pages}{121--128}
	(\bibinfo{year}{2017}).
	
	\bibitem{105_Felsenstein2008-gk}
	\bibinfo{author}{Felsenstein, J.}
	\newblock \bibinfo{title}{Bootstrap confidence levels for phylogenetic trees}.
	\newblock In \bibinfo{editor}{Morris, C.~N.} \& \bibinfo{editor}{Tibshirani,
		R.} (eds.) \emph{\bibinfo{booktitle}{The Science of Bradley Efron}}, Springer
	Series in Statistics, \bibinfo{pages}{336--343} (\bibinfo{publisher}{Springer
		New York}, \bibinfo{address}{New York, NY}, \bibinfo{year}{2008}).
	
	\bibitem{106_Powell2012-hb}
	\bibinfo{author}{Powell, W.~B.} \emph{et~al.}
	\newblock \emph{\bibinfo{title}{Optimal learning}} (\bibinfo{publisher}{Wiley},
	\bibinfo{address}{Oxford}, \bibinfo{year}{2012}).
	
	\bibitem{107_Powell2010-cp}
	\bibinfo{author}{Powell, W.~B.} \emph{et~al.}
	\newblock \bibinfo{title}{The knowledge gradient for optimal learning}.
	\newblock In \emph{\bibinfo{booktitle}{Wiley Encyclopedia of Operations
			Research and Management Science}} (\bibinfo{publisher}{John Wiley \& Sons,
		Inc.}, \bibinfo{year}{2010}).
	
	\bibitem{108_Ryzhov2012-aw}
	\bibinfo{author}{Ryzhov, I.~O.}, \bibinfo{author}{Powell, W.~B.} \&
	\bibinfo{author}{Frazier, P.~I.}
	\newblock \bibinfo{title}{The knowledge gradient algorithm for a general class
		of online learning problems}.
	\newblock \emph{\bibinfo{journal}{Oper. Res.}} \textbf{\bibinfo{volume}{60}},
	\bibinfo{pages}{180--195} (\bibinfo{year}{2012}).
	
	\bibitem{109_Micchelli2005-lj}
	\bibinfo{author}{Micchelli, C.~A.} \& \bibinfo{author}{Pontil, M.}
	\newblock \bibinfo{title}{On learning vector-valued functions}.
	\newblock \emph{\bibinfo{journal}{Neural Comput.}}
	\textbf{\bibinfo{volume}{17}}, \bibinfo{pages}{177--204}
	(\bibinfo{year}{2005}).
	
	\bibitem{110_Alvarez2012-wa}
	\bibinfo{author}{{\'A}lvarez, M.~A.}, \bibinfo{author}{Rosasco, L.} \&
	\bibinfo{author}{Lawrence, N.~D.}
	\newblock \emph{\bibinfo{title}{Kernels for Vector-valued Functions: A Review}}
	(\bibinfo{publisher}{Now Publishers Incorporated}, \bibinfo{year}{2012}).
	
	\bibitem{111_Forrester2007-so}
	\bibinfo{author}{Forrester, A. I.~J.}, \bibinfo{author}{S{\'o}bester, A.} \&
	\bibinfo{author}{Keane, A.~J.}
	\newblock \bibinfo{title}{Multi-fidelity optimization via surrogate modelling}.
	\newblock \emph{\bibinfo{journal}{Proc. R. Soc. A}}
	\textbf{\bibinfo{volume}{463}}, \bibinfo{pages}{3251--3269}
	(\bibinfo{year}{2007}).
	
	\bibitem{112_Perdikaris2015-yt}
	\bibinfo{author}{Perdikaris, P.}, \bibinfo{author}{Venturi, D.},
	\bibinfo{author}{Royset, J.~O.} \& \bibinfo{author}{Karniadakis, G.~E.}
	\newblock \bibinfo{title}{Multi-fidelity modelling via recursive co-kriging and
		{Gaussian-Markov} random fields}.
	\newblock \emph{\bibinfo{journal}{Proc. Math. Phys. Eng. Sci.}}
	\textbf{\bibinfo{volume}{471}}, \bibinfo{pages}{20150018}
	(\bibinfo{year}{2015}).
	
	\bibitem{113_Dudiy2006-ge}
	\bibinfo{author}{Dudiy, S.~V.} \& \bibinfo{author}{Zunger, A.}
	\newblock \bibinfo{title}{Searching for alloy configurations with target
		physical properties: impurity design via a genetic algorithm inverse band
		structure approach}.
	\newblock \emph{\bibinfo{journal}{Phys. Rev. Lett.}}
	\textbf{\bibinfo{volume}{97}}, \bibinfo{pages}{046401}
	(\bibinfo{year}{2006}).
	
\end{thebibliography}

\end{document}